\documentclass[aps,preprint,showpacs,showkeys,superscriptaddress,groupedaddress]{revtex4}  

\usepackage{graphicx}  
\usepackage{float}
\usepackage{dcolumn}   
\usepackage{bm}        
\usepackage{amssymb}   
\usepackage{slashed}   
\usepackage{amsmath}   
\usepackage{simplewick} 
\usepackage{verbatim}  
\usepackage{color} 
\usepackage{appendix} 
\usepackage{subfig}

\usepackage[bookmarks=true,colorlinks=true,linkcolor=blue,unicode=true]{hyperref}

\begin{document}

\widetext


\title{Effective field theory of the Higgs Mode in a two dimensional dilute Bose gas }

%
%
\author{Ji-Chong Yang}

\affiliation{Department of Physics  \&  State Key Laboratory of Surface Physics,   Fudan University,\\ Shanghai 200433, China}

\affiliation{ Department of Physics, Liaoning Normal University, Dalian 116029, China}

\author{Yu Shi}
\email[]{yushi@fudan.edu.cn}
\thanks{Corresponding author}

\affiliation{Department of Physics  \&  State Key Laboratory of Surface Physics,   Fudan University,\\ Shanghai 200433, China}

%
%
%
\vskip 0.25cm


\begin{abstract}
We investigate the spectral function of the Higgs mode in a two dimensional  Bose gas, by using the effective field theory in the zero temperature limit. Our approach explains  the experimental feature  that the peak of the spectral function is a soft continuum  rather than a sharp peak,  broadened  and vanishing   in  the superfluid phase, which   cannot be explained in terms of the $O(2)$ model. We also find that  the scalar susceptibility is the same as the longitudinal susceptibility.
\end{abstract}



\maketitle

\section{\label{sec:1}Introduction}

In condensed matter physics, the   Higgs mode was first explored  in superconductivity~\cite{littlewood}, and  has recently been observed in ultra-cold Bosons  in both three and two dimensional optical lattices~\cite{3DOpticalLattice,2DOpticalLattice}.  Higgs modes in various other systems have also been  studied~\cite{OtherExperiments}. The Higgs mode  in Boson systems in an optical lattice   has been theoretically  studied by using $O(2)$ model~\cite{O2Model1,Podolsky1,Podolsky2,O22loop,3DO21,3DO22}. However, it was found experimentally that  the dependence of the response   function on  the frequency exhibits a broad continuum rather than a sharp peak~\cite{2DOpticalLattice,3DO22}. This feature cannot be explained in terms of the $O(2)$ model~\cite{Podolsky1,Podolsky2,O22loop}. Moreover,  as the system deviates from the critical point and enters the superfluid phase, the response function broadens and vanishes~\cite{2DOpticalLattice}. This phenomenon  also exists in Fermi superfluid~\cite{FermionSuperfluid}.

In this paper, we study the spectral function of the Higgs mode in $2+1$ dimensions by  using the effective field theory~(EFT) in the  zero temperature limit~\cite{EFT1,EFT2,EFT3}. We study the spectral functions of both the longitudinal and the  scalar susceptibilities, which are found to be the same when ${\bf q}=0$. We also find that the feature of the  peak of the spectral function is consistent with the experiment. The peak of the longitudinal susceptibility  is   soft, as observed in  the  experiment. Furthermore, our theory reproduces the disappearance of the  Higgs mode in  the ordered phase, as observed in  the  experiment.

The rest of the paper is organized as the following. In Sec.~\ref{sec:2}, we briefly introduce the EFT  in a two dimensional Bose gas. The correlation functions of the Higgs mode are calculated in Sec.~\ref{sec:3}. In Sec.~\ref{sec:4}, we present the numerical study. Sec.~\ref{sec:5} is a summary.

\section{\label{sec:2}Effective field theory}

In imaginary time representation,  the action of EFT   can be written  as~\cite{EFT3}
\begin{equation}
\begin{split}
&S[\psi ^*,\psi]=\int _0^{\beta} d\tau \int d^D x \left\{\psi ^*\left[\frac{\partial}{\partial \tau}-\nabla^2-r\right]\psi+\frac{1}{2}g(\psi ^*\psi)^2\right.\\
&\left.+\frac{1}{2}h[\nabla(\psi^*\psi)]^2+\frac{g_3}{36}(\psi^*\psi)^3
+\ldots\right\},
\end{split}
\label{eq.2.1}
\end{equation}
where $r=\mu$ is the chemical potential, $g$, $h$, and $g_3$ are coupling constants that  can be determined from fitting the results with the experimental data, ``$\ldots$''  represents the higher order terms. We consider only the leading order, that is, the case that $g$ is the only nonzero coupling constant.

One can compare the EFT with the $O(2)$ model, which is a relativistic model and can describe the Bose gas at the vicinity of the  critical point, with the action
\begin{equation}
\begin{split}
&S[\Phi]=\int d^{D+1} x \left\{\frac{1}{2}\left(\partial _{\mu} \Phi\right)^2 -\frac{m^2}{2}\Phi ^2 + \frac{U}{4}\Phi ^2\Phi ^2\right\},
\end{split}
\label{eq.2.2}
\end{equation}
where $\Phi$ is a two component vector. In EFT, if we write $\psi$ as a two component vector, the only difference with the $O(2)$ model is the derivative with respect to $t$. Similar to the parametrization of $O(2)$ model,   $\psi$ can be parameterized  as
\begin{equation}
\begin{split}
&\psi =v+\frac{1}{\sqrt{2}}\left(\psi _1+i\psi _2\right).\\
\end{split}
\label{eq.2.3}
\end{equation}
Then the action of effective field theory can be  written as
\begin{equation}
\begin{split}
&S[v,\psi _1,\psi _2]=S_v[v]+S_{\rm free}[v,\psi _1,\psi _2]+S_{\rm int}[v, \psi _1, \psi _2],\\
&S_v[v]=\int d\tau \int d^d x \left[-r v^2+\frac{1}{2}gv^4\right],\\
&S_{\rm free}[v,\psi _1,\psi _2]=\int d\tau \int d^d x \left[\frac{i}{2}\left(\psi _1\dot{\psi _2}-\dot{\psi }_1\psi _2\right)+\frac{1}{2}\psi _1(-\nabla ^2+X)\psi _1+\frac{1}{2}\psi _2(-\nabla ^2+Y)\psi _2\right],\\
&S_{\rm int}[v,\psi _1,\psi _2]=\int d\tau \int d^d x \left[-\sqrt{2}(\mu-gv^2)v\psi _1+\frac{gv}{\sqrt{2}}\psi _1 (\psi _1^2+\psi _2^2)+\frac{1}{8}g(\psi _1^2+\psi _2^2)^2\right],\\
&X=-r+3gv^2,\;Y=-r +gv^2.\\
\end{split}
\label{eq.2.4}
\end{equation}

\subsection{\label{sec:2.1}Feynman rules}

The lowest-energy classical configuration of the potential is a constant field $\psi = v$, with
\begin{equation}
\begin{split}
&v=\sqrt{\frac{r}{g}}.
\end{split}
\label{eq.2.5}
\end{equation}
By choosing such a minimum, the global $U(1)$ gauge symmetry is spontaneously  broken.

There are ultra violet  (UV) divergences  in 1-loop calculation in $1+2$ dimensions. One can deal with the UV divergences by renormalization, i.e.  rescaling the field as $\psi \to Z^{\frac{1}{2}}\psi$ and introducing  the counter terms   defined as~\cite{PeskinBook}
\begin{equation}
\begin{split}
&\delta _z=Z-1,\;\;\delta _{r}=r _0Z-r,\;\;\delta _g=g_0Z-g.\\
\end{split}
\label{eq.2.6}
\end{equation}
where $r _0$ and $g_0$ are bare chemical potential and bare coupling constant, respectively,  to replace $r$ and $g$ in the original Lagrangian. We find that $\delta _g$ is sufficient to cancel the UV divergence showing up in 1-loop calculation. Thus we use $Z=1$, $\delta _z=\delta _{r}=0$. With only one counter term, we need only one renormalization condition. Similar to the $O(2)$ model, we use the renormalization condition~\cite{Podolsky1,O22loop,PeskinBook}
\begin{equation}
\begin{split}
&\langle \psi _1 \rangle=0.\\
\end{split}
\label{eq.2.7}
\end{equation}
The action at classical minimum with counter terms can be written as
\begin{equation}
\begin{split}
&S[v,\psi _1,\psi _2]=S_v[v]+S_{\rm free}[v,\psi _1,\psi _2]+S_{\rm int}[v, \psi _1, \psi _2]+S_{c}[v, \psi _1, \psi _2]\\
&S_v[v]=\int d\tau \int d^d x \left[\frac{1}{2}(\delta _g-g)v^4\right],\\
&S_{\rm free}[v,\psi _1,\psi _2]=\int d\tau \int d^d x \left[\frac{i}{2}\left(\psi _1\dot{\psi} _2-\dot{\psi} _1\psi _2\right)-\frac{1}{2}\psi _1(\nabla ^2-2gv^2)\psi _1-\frac{1}{2}\psi _2(\nabla ^2)\psi _2\right],\\
&S_{\rm int}[v,\psi _1,\psi _2]=\int d\tau \int d^d x \left[\frac{gv}{\sqrt{2}}\psi _1 (\psi _1^2+\psi _2^2)+\frac{1}{8}g(\psi _1^2+\psi _2^2)^2\right],\\
&S_{c}[v, \psi _1, \psi _2]=\int d\tau \int d^d x \left[\frac{3}{2}\delta _gv^2\psi _1^2+\frac{1}{2}\delta _gv^2\psi _2^2+\sqrt{2}\delta _gv^3\psi _1+\frac{\delta _gv}{\sqrt{2}}\psi _1(\psi _1^2+\psi _2^2)+\frac{\delta _g}{8}(\psi _1^2+\psi _2^2)^2\right].\\
\end{split}
\label{eq.2.8}
\end{equation}

\begin{figure}
\includegraphics[scale=0.7]{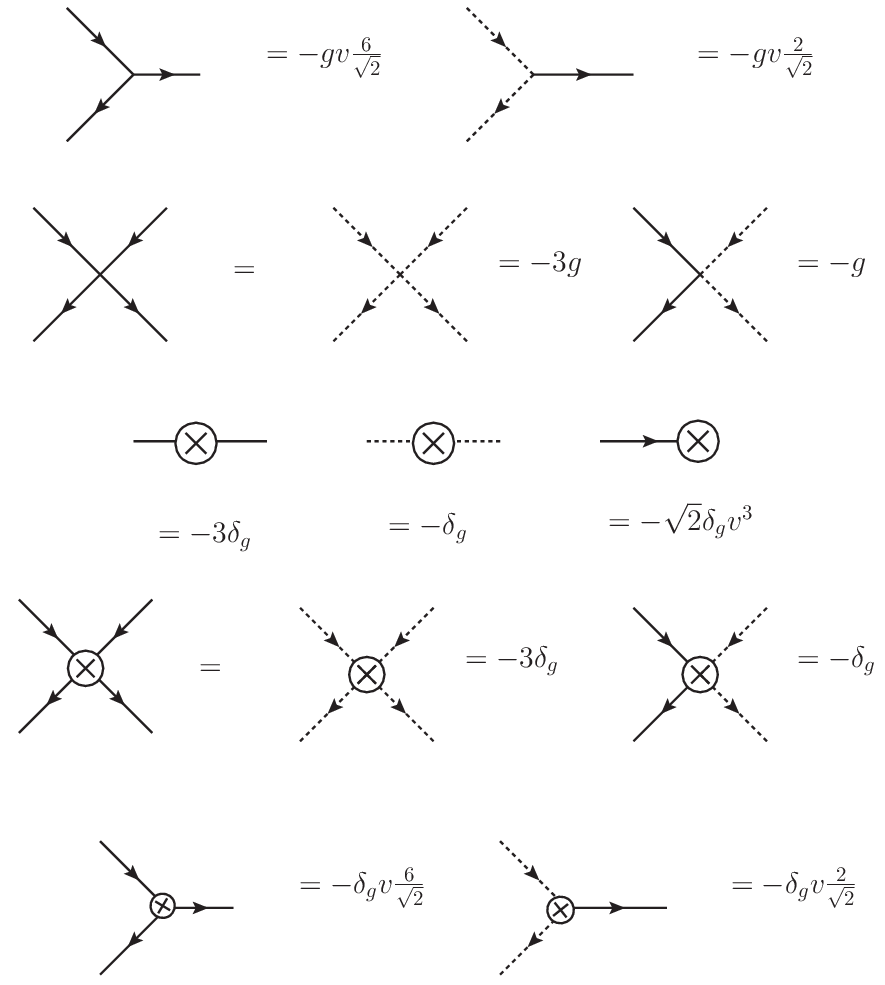}
\caption{Feynman rules of EFT. The solid line is the propagator of $\psi _1$, the dotted line is the propagator of $\psi _2$.}
\label{fig:rules}
\end{figure}

The propagator can be written as~\cite{EFT3}
\begin{equation}
\begin{split}
&D(\omega,{\bf p})=\frac{1}{\omega^2 +\epsilon ^2(p)}\left(\begin{array}{cc}p^2&\omega\\ -\omega &p^2+2gv^2\end{array}\right),\\
&\epsilon ({\bf p})=\sqrt{p^2(p^2+2gv^2)},\\
\end{split}
\label{eq.2.9}
\end{equation}
where we have  used a Nambu spinor to denote $\psi _1$ and $\psi _2$. The Feynman rules for the vertices are  shown in Fig.~\ref{fig:rules}.

\subsection{\label{sec:2.2}Susceptibility}

The observable we are interested in is the spectral function of Higgs  mode. The spectral function can be defined via dynamic susceptibility as~\cite{Podolsky1,CondensedMatterBook}
\begin{equation}
\begin{split}
&\chi_{AB}''({\bf q}, \omega)={\rm Im}(\chi _{AB}({\bf q}, i\omega\to \omega +i 0^+)),
\end{split}
\label{eq.2.10}
\end{equation}
so that   $\chi_{AB}''({\bf q}, \omega)$ are the imaginary parts of retarded correlation functions, which can be obtained from thermal correlation functions $\chi _{AB}({\bf q}, i\omega)$ by analytical continuation  $i\omega \to \omega +i0^+$. The thermal correlation function $\chi _{AB}({\bf q}, i\omega)$ can be calculated in imaginary time representation.

The scalar susceptibility is introduced in Ref.~\cite{Podolsky1}. It was argued that  to observe the Higgs mode in experiments, one should try to measure the spectral function of the scalar susceptibility. The scalar susceptibility can be associated with the parameterization~\cite{EFT3}
\begin{equation}
\begin{split}
&\psi(x,t) = \sqrt{n(x,t)}e^{i\phi (x,t)},\;\;n(x,t)=v^2+\rho(x,t).\\
\end{split}
\label{eq.2.11}
\end{equation}
Using Eq.~(\ref{eq.2.3}), we find
\begin{equation}
\begin{split}
&\rho(x,t)=\sqrt{2}v\psi _1+\frac{1}{2}\psi _1^2+\frac{1}{2}\psi _2^2.  \\
\end{split}
\label{eq.2.12}
\end{equation}
Similar to the approach in Ref.~\cite{Podolsky1},  it is found that
\begin{equation}
\begin{split}
&\chi _{\rho\rho}=2v^2\chi _{\psi _1\psi _1}+\sqrt{2}v\left(\chi _{\psi _1\psi _1^2}+\chi _{\psi _1\psi _2^2}\right)+\frac{1}{4}\left(\chi _{\psi _1^2\psi _1^2}+\chi _{\psi _2^2\psi _2^2}+2\chi _{\psi _1^2\psi _2^2}\right).
\end{split}
\label{eq.2.13}
\end{equation}

In this paper, we study the spectral functions of both longitudinal susceptibility $\chi ''_{\psi_1\psi_1}$ and scalar susceptibility $\chi ''_{\rho\rho}$.

\section{\label{sec:3}Calculation of correlation functions}

Throughout this paper, we consider zero temperature limit and  $2+1$ dimensions. We use dimensional regulation~(DR)~\cite{DR} to regulate the UV divergence. For simplicity, in $D=2-\epsilon$ dimensions, we define $N_{\rm UV}$ as
\begin{equation}
\begin{split}
&N_{\rm UV}\equiv \frac{2}{\epsilon}-\gamma _E+\log (16\pi)+\log \frac{M^2}{2gv^2},
\end{split}
\label{eq.3.1}
\end{equation}
where $\gamma _E$ is the Euler constant, $M$ is renormalization scale.

\subsection{\label{sec:3.1}1-loop level}

\subsubsection{\label{sec:3.1.1}Counter terms at 1-loop order}

\begin{figure}
\includegraphics[scale=0.6]{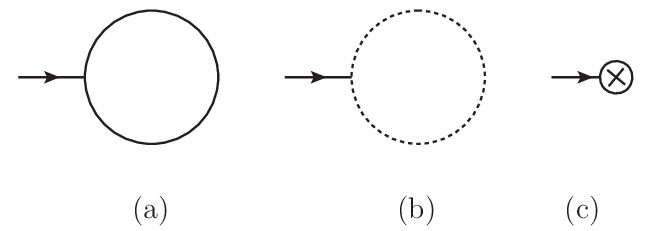}
\caption{The diagrams of 1PI contribution $\langle \psi _1 \rangle$ at 1-loop level.}
\label{fig:tadpole}
\end{figure}

The renormalization condition in Eq.~(\ref{eq.2.7}) requires the 1-particle-irreducible~(1PI) tadpole diagrams of $\psi _1$ vanish. All the 1PI diagrams at 1-loop level are shown in Fig.~\ref{fig:tadpole}. The diagrams shown in Fig.~\ref{fig:tadpole}.~(a), (b) and (c) are denoted as $I_a^t$, $I_b^t$ and $I_c^t$ respectively, and can be written as

\begin{equation}
\begin{split}
&I_a^t=-gv\frac{6}{\sqrt{2}}f_a^t,\;\;\;\;I_b^t=-gv\frac{2}{\sqrt{2}}f_b^t,\;\;\;\;I_c^t=-\sqrt{2}\delta _g^{(1)} v^3,\\
\end{split}
\label{eq.3.2}
\end{equation}
where we use the superscript of  $\delta _g^{(1)}$ to denote $\delta _g$ at 1-loop level, $f_a^t$ and $f_b^t$ are obtained in Eq.~(\ref{eq.a.17}). Using the renormalization condition
\begin{equation}
\begin{split}
&\langle \psi _1\rangle =I_a^t+I_b^t+I_c^t=0,\\
\end{split}
\label{eq.3.3}
\end{equation}
we find that in $D=2-\epsilon$ dimensions,  the counter term at 1-loop level can be written as
\begin{equation}
\begin{split}
&\delta _g^{(1)}=\frac{g^2}{8\pi}\left(N_{\rm UV}-2\right),\\
\end{split}
\label{eq.3.4}
\end{equation}
where $N_{\rm UV}$ is defined in Eq.~(\ref{eq.3.1}).

\subsubsection{\label{sec:3.1.2}  1PI contribution to self-energy at 1-loop order}

\begin{figure}
\includegraphics[scale=0.8]{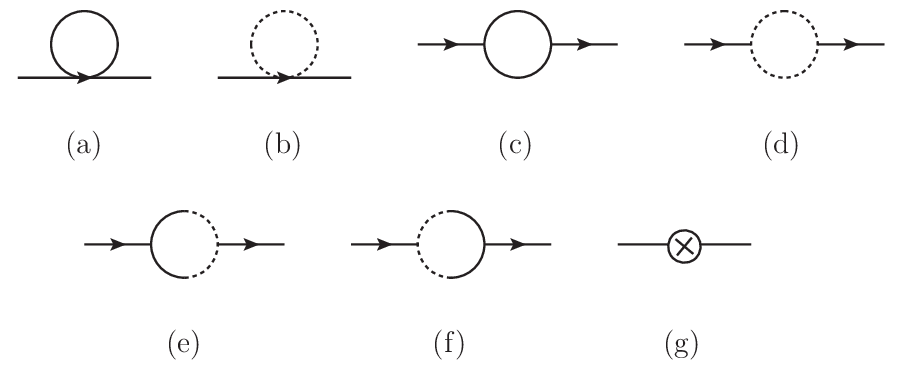}
\caption{The diagrams of 1PI contribution to $\Pi _{11}$ at 1-loop level.}
\label{fig:psi1loop}
\end{figure}

The 1PI contribution to self-energy of $\psi _1$ is denoted as $\Pi _{11}$.  The diagrams contributing  to $\Pi _{11}$ at 1-loop level are shown in Fig.~\ref{fig:psi1loop}. The diagrams shown in Fig.~\ref{fig:psi1loop}.~(a), (b), (c), (d), (e), (f) and (g) are denoted as $I_a^{\psi _1}$, $I_b^{\psi _1}$, $I_c^{\psi _1}$, $I_d^{\psi _1}$, $I_e^{\psi _1}$, $I_f^{\psi _1}$ and $I_g^{\psi _1}$, respectively, and can be written as
\begin{equation}
\begin{split}
&I_a^{\psi _1}=-3gf_a^t,\;\;\;\;I_b^{\psi _1}=-gf_b^t,\;\;\;\;I_c^{\psi _1}(\omega _q,q^2)=18g^2v^2f_a^p(\omega _q,q^2),\\
&I_d^{\psi _1}(\omega _q,q^2)=2g^2v^2f_b^p(\omega _q,q^2),\;\;I_e^{\psi _1}(\omega _q,q^2)=I_f^{\psi _1}(\omega _q,q^2)=6g^2v^2f_c^p(\omega _q,q^2),\;\;I_g^{\psi _1}=-3\delta _g,\\
\end{split}
\label{eq.3.5}
\end{equation}
where $f_a^t$, $f_b^t$, $f_a^p(q^2)$, $f_b^p(q^2)$ and $f_c^p(q^2)$ are given in Eqs.~(\ref{eq.a.17}), (\ref{eq.a.34}), (\ref{eq.a.35}) and (\ref{eq.a.28}). $\delta _g$ are given in Eq.~(\ref{eq.3.4}). We find
\begin{equation}
\begin{split}
&\Pi_{11}(\omega_q, q^2)=\sum _{n=a,\ldots,g}I_n^{\psi _1}=\frac{g^2 v^2 \left(-\frac{\left(\omega_q^2-20 g^2 v^4\right) \sec ^{-1}\left(\frac{2 g v^2}{\omega_q}\right)}{\sqrt{4 g^2 v^4-\omega_q^2}}-4 \pi  g v^2+2 \omega_q\right)}{4 \pi  \omega_q}\\
&-\frac{g^3 q^2 v^4}{4 \pi  \omega_q^3 \left(4 g^2 v^4-\omega_q^2\right)^{3/2}} \left(\sqrt{4 g^2 v^4-\omega_q^2} \left(104 \pi  g^3 v^6-100 g^2 \omega_q v^4-26 \pi  g \omega_q^2 v^2+21 \omega_q^3\right)\right.\\
&\left.-4 \left(100 g^4 v^8-37 g^2 \omega_q^2 v^4+2 \omega_q^4\right) \sec ^{-1}\left(\frac{2 g v^2}{\omega_q}\right)\right)+\mathcal{O}(q^4).
\end{split}
\label{eq.3.6}
\end{equation}

\begin{figure}
\includegraphics[scale=0.7]{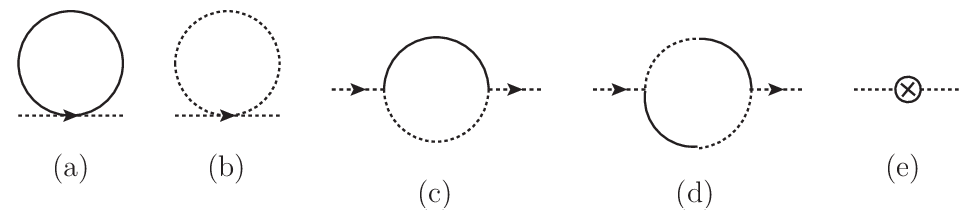}
\caption{The diagrams of 1PI contribution to $\Pi _{22}$ at 1-loop level.}
\label{fig:psi2loop}
\end{figure}

The 1PI contribution to self-energy of $\psi _2$ is denoted as $\Pi _{22}$,  The diagrams contributing to $\Pi _{22}$ at 1-loop level are shown in Fig.~\ref{fig:psi2loop}. The diagrams shown in Fig.~\ref{fig:psi2loop}.~(a), (b), (c), (d) and (e) are denoted as $I_a^{\psi _2}$, $I_b^{\psi _2}$, $I_c^{\psi _2}$, $I_d^{\psi _2}$, and $I_e^{\psi _2}$ respectively, and can be written as
\begin{equation}
\begin{split}
&I_a^{\psi _2}=-gf_a^t,\;\;\;\;I_b^{\psi _2}=-3gf_b^t,\;\;\;\;I_c^{\psi _2}(\omega_q,q^2)=2g^2v^2f_d^p(\omega_q,q^2),\\
&I_d^{\psi _2}(\omega_q,q^2)=2g^2v^2f_e^p(\omega_q,q^2),\;\;\;\;I_e^{\psi _2}=-\delta _gv^2,\\
\end{split}
\label{eq.3.7}
\end{equation}
where $f_e^p(q^2)=-2f_c^p(q^2)$, $f_a^t$, $f_b^t$, $f_c^p(q^2)$ and $f_d^p(q^2)$ are given in Eqs.~(\ref{eq.a.17}), (\ref{eq.a.28}) and (\ref{eq.a.36}), while  $\delta _g$ is given in Eq.~(\ref{eq.3.4}). We find
\begin{equation}
\begin{split}
&\Pi _{22}(\omega_q, q^2)=\sum _{n=a,\ldots,e}I_n^{\psi _2}=-\frac{g^2 \omega_q v^2 \sec ^{-1}\left(\frac{2 g v^2}{\omega_q}\right)}{4 \pi  \sqrt{4 g^2 v^4-\omega_q^2}}\\
&+\frac{g^3 q^2 v^4 \left(4 g^2 v^4 \sqrt{4 g^2 v^4-\omega_q^2} \sec ^{-1}\left(\frac{2 g v^2}{\omega_q}\right)-4 g^2 \omega_q v^4+\omega_q^3\right)}{4 \pi  \omega_q \left(\omega_q^2-4 g^2 v^4\right)^2}+\mathcal{O}(q^4).\\
\end{split}
\label{eq.3.8}
\end{equation}

\begin{figure}
\includegraphics[scale=0.6]{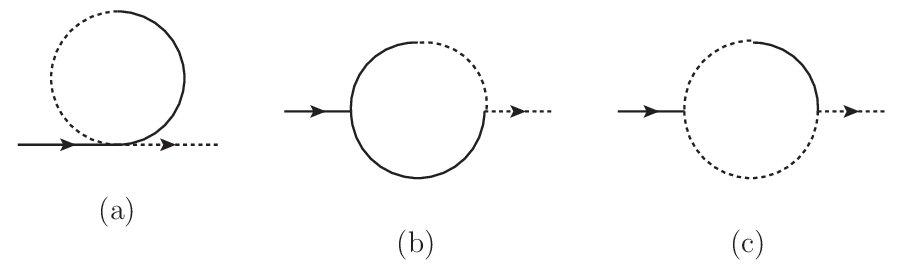}
\caption{The diagrams of 1PI contribution to $\Pi _{12}$ at 1-loop level.}
\label{fig:psi12loop}
\end{figure}

The 1PI contribution to self-energy that one $\psi _1$ is annihilated while a $\psi _2$ is created is denoted as $\Pi _{12}$.  The diagrams contributing  to $\Pi _{12}$ at 1-loop level are shown in Fig.~\ref{fig:psi12loop}. The diagrams shown in Fig.~\ref{fig:psi12loop}.~(a) (b) and (c) are denoted as $I_a^{\psi _1\psi _2}$, $I_b^{\psi _1\psi _2}$ and $I_c^{\psi _1\psi _2}$ respectively, and can be written as
\begin{equation}
\begin{split}
&I_a^{\psi _1\psi _2}=0,\;\;\;\;I_b^{\psi _1\psi _2}(\omega_q,q^2)=6g^2v^2f_f^p(\omega_q,q^2),\;\;\;\;I_c^{\psi _1\psi _2}(\omega_q,q^2)=-2g^2v^2f_g^p(\omega_q,q^2),\\
\end{split}
\label{eq.3.9}
\end{equation}
where $f_f^p(q^2)$ and $f_p^p(q^2)$ are given in Eqs.~(\ref{eq.a.37}) and (\ref{eq.a.38}). We find
\begin{equation}
\begin{split}
&\Pi _{12}(\omega _q,q^2)=I_b^{\psi _1\psi _2}(\omega_q,q^2)+I_c^{\psi _1\psi _2}(\omega_q,q^2)=\frac{g^3 v^4 \sec ^{-1}\left(\frac{2 g v^2}{\omega _q}\right)}{\pi  \sqrt{4 g^2 v^4-\omega _q^2}}-\frac{g^2 v^2}{8}\\
&-\frac{g^2 q^2 v^2}{8 \pi  \omega _q^2 \left(4 g^2 v^4-\omega _q^2\right)^{3/2}} \left(\sqrt{4 g^2 v^4-\omega _q^2} \left(32 \pi  g^3 v^6-28 g^2 \omega _q v^4-8 \pi  g \omega _q^2 v^2+5 \omega _q^3\right)\right.\\
&\left.-4 \left(28 g^4 v^8-13 g^2 \omega _q^2 v^4+\omega _q^4\right) \sec ^{-1}\left(\frac{2 g v^2}{\omega _q}\right)\right)+\mathcal{O}(q^4).
\end{split}
\label{eq.3.10}
\end{equation}

\subsubsection{\label{sec:3.1.3} 1PI contribution to cross susceptibilities}

\begin{figure}
\includegraphics[scale=0.6]{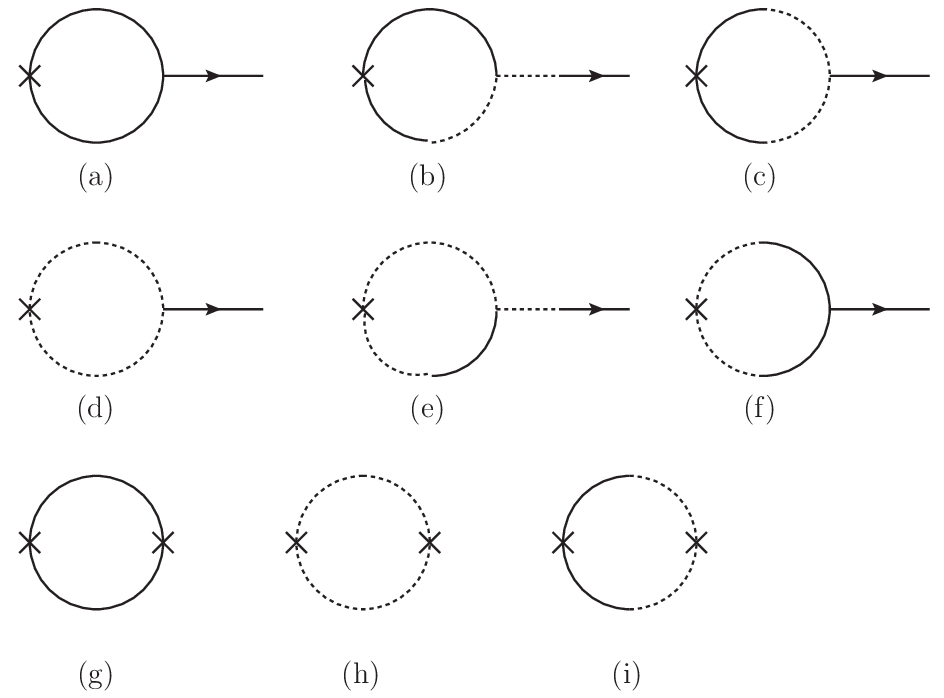}
\caption{The diagrams of 1PI contribution to cross-susceptibilities at 1-loop level.}
\label{fig:crossloop}
\end{figure}

The cross susceptibilities at 1-loop level are denoted as $\chi _{\psi _1^2\psi _1}^{(1)}$, $\chi _{\psi _2^2\psi _1}^{(1)}$, $\chi _{\psi _1^2\psi _1^2}^{(1)}$, $\chi _{\psi _2^2\psi _2^2}^{(1)}$ and $\chi _{\psi _1^2\psi _2^2}^{(1)}$, where we use the superscript to denote the susceptibilities at 1-loop level.  The 1PI diagrams contributing  to cross susceptibilities at 1-loop level are shown in Fig.~\ref{fig:crossloop}. The diagrams shown in Fig.~\ref{fig:crossloop}.~(a), (b), (c), (d), (e), (f), (g), (h) and (i) are denoted as $I_a^{cs}$, $I_b^{cs}$, $I_c^{cs}$, $I_d^{cs}$, $I_e^{cs}$, $I_f^{cs}$, $I_g^{cs}$, $I_h^{cs}$ and $I_i^{cs}$ respectively, and can be written as
\begin{equation}
\begin{split}
&I_a^{ct}(\omega_q,q^2)=-2\frac{6gv}{\sqrt{2}}f_a^p(\omega_q,q^2)\frac{q^2}{\omega_q^2+\epsilon^2 (q)},\;
 I_b^{ct}(\omega_q,q^2)=-2\frac{2gv}{\sqrt{2}}f_f^p(\omega_q,q^2)\frac{\omega_q}{\omega_q^2+\epsilon^2 (q)},\\
&I_c^{ct}(\omega_q,q^2)=-2\frac{2gv}{\sqrt{2}}f_c^p(\omega_q,q^2)\frac{q^2}{\omega_q^2+\epsilon^2 (q)},\;
 I_d^{ct}(\omega_q,q^2)=-2\frac{2gv}{\sqrt{2}}f_b^p(\omega_q,q^2)\frac{q^2}{\omega_q^2+\epsilon^2 (q)},\\
&I_e^{ct}(\omega_q,q^2)=2\frac{2gv}{\sqrt{2}}f_g^p(\omega_q,q^2)\frac{\omega_q}{\omega_q^2+\epsilon^2 (q)},\;
 I_f^{ct}(\omega_q,q^2)=-2\frac{6gv}{\sqrt{2}}f_c^p(\omega_q,q^2)\frac{q^2}{\omega_q^2+\epsilon^2 (q)},\\
&I_g^{ct}(\omega_q,q^2)=4f_a^p(\omega_q,q^2),\;\;\;\;I_h^{ct}(\omega_q,q^2)=4f_b^p(\omega_q,q^2),\;\;\;\;I_i^{ct}(\omega_q,q^2)=4f_c^p(\omega_q,q^2).\\
\end{split}
\label{eq.3.11}
\end{equation}
We find at 1-loop level that
\begin{equation}
\begin{split}
&\chi _{\psi _1^2\psi _1}^{(1)}(\omega_q,{\bf q})+\chi _{\psi _2^2\psi _1}^{(1)}(\omega_q,{\bf q})=\sum _{n=a,\ldots,f}I_n^{ct}(\omega_q,q^2)\\
&=-\frac{g^2 v^3 \sec ^{-1}\left(\frac{2 g v^2}{\omega _q}\right)}{\pi  \omega _q \sqrt{8 g^2 v^4-2 \omega _q^2}}+\frac{q^2g^2 v^3}{2 \pi  \omega _q^3 \sqrt{8 g^2 v^4-2 \omega _q^2} \left(2 g q^2 v^2+q^4+\omega _q^2\right) \left(\omega _q^2-4 g^2 v^4\right)}\\
&\times\left\{4 \left(\omega _q^2-3 g^2 v^4\right) \left(\omega _q^2 \left(q^2-2 g v^2\right)-12 g^2 q^2 v^4\right) \sec ^{-1}\left(\frac{2 g v^2}{\omega _q}\right)+\sqrt{4 g^2 v^4-\omega _q^2} \left[-40 \pi  g^3 q^2 v^6\right.\right.\\
&\left.\left.+4 g^2 v^4 \omega _q \left(9 q^2-2 \pi  \omega _q\right)+2 g v^2 \omega _q^2 \left(5 \pi  q^2+\omega _q\right)+\omega _q^3 \left(2 \pi  \omega _q-7 q^2\right)\right]\right\}+\mathcal{O}(q^4),\\
&\chi _{\psi _1^2\psi _1^2}^{(1)}(\omega_q,{\bf q})+\chi _{\psi _2^2\psi _2^2}^{(1)}(\omega_q,{\bf q})+2\chi _{\psi _1^2\psi _2^2}^{(1)}(\omega_q,{\bf q})=I_g^{ct}(\omega_q,q^2)+I_h^{ct}(\omega_q,q^2)+2I_i^{ct}(\omega_q,q^2)\\
&=\frac{2 g^2 v^4 \sec ^{-1}\left(\frac{2 g v^2}{\omega _q}\right)}{\pi  \omega _q \sqrt{4 g^2 v^4-\omega _q^2}}-\frac{g q^2 v^2}{\pi  \omega_q^3 \left(4 g^2 v^4-\omega_q^2\right)^{3/2}} \left(8 g^2 v^4 \left(\omega_q^2-3 g^2 v^4\right) \sec ^{-1}\left(\frac{2 g v^2}{\omega_q}\right)\right.\\
&\left.+\left(8 \pi  g^3 v^6-6 g^2 \omega_q v^4-2 \pi  g \omega_q^2 v^2+\omega_q^3\right) \sqrt{4 g^2 v^4-\omega_q^2}\right)+\mathcal{O}(q^4).
\end{split}
\label{eq.3.12}
\end{equation}

At 1-loop level, the self energy can be written as
\begin{equation}
\begin{split}
&\Sigma ^{(1)}(\omega_q, q^2)=D_0(\omega _q, {\bf q})+D_0(\omega _q, {\bf q})\cdot \Pi \cdot D_0(\omega _q, {\bf q})
\end{split}
\label{eq.3.13}
\end{equation}
where $D(\omega_q, q^2)$ is defined in Eq.~(\ref{eq.2.9}), while  $\Pi(\omega_q, q^2)$ is defined as
\begin{equation}
\begin{split}
&\Pi(\omega_q, q^2)\equiv \left(\begin{array}{cc}\Pi _{11}(\omega_q,q^2)&-\Pi _{12}(\omega_q,q^2)\\ \Pi _{12}(\omega_q,q^2)&\Pi _{22}(\omega_q,q^2)\end{array}\right),\\
\end{split}
\label{eq.3.14}
\end{equation}

The thermal correlation function $\chi^{(1)} _{\psi_1\psi_1}$ at 1-loop level is the matrix element $\left(\Sigma^{(1)}(\omega_q, q^2)\right)^{11}$ at 1-loop level. One can find that there is  an infrared singularity in $\chi _{\psi_1\psi_1}$ when $\omega _q\to 0$ and ${\bf q}=0$. However, for scalar susceptibility, such an infrared singularity is cancelled, as in the $O(2)$ model~\cite{Podolsky1}. Using Eq.~(\ref{eq.2.13}), we find that
\begin{equation}
\begin{split}
&\chi^{(1)} _{\rho\rho}(\omega_q,{\bf q})=\frac{q^2 \left(g v^2+8 \pi  v^2\right)}{4 \pi  \omega_q^2}+\mathcal{O}(q^4)
\end{split}
\label{eq.3.15}
\end{equation}

\subsection{\label{sec:3.3}Higher order contributions}

\begin{figure}
\includegraphics[scale=0.6]{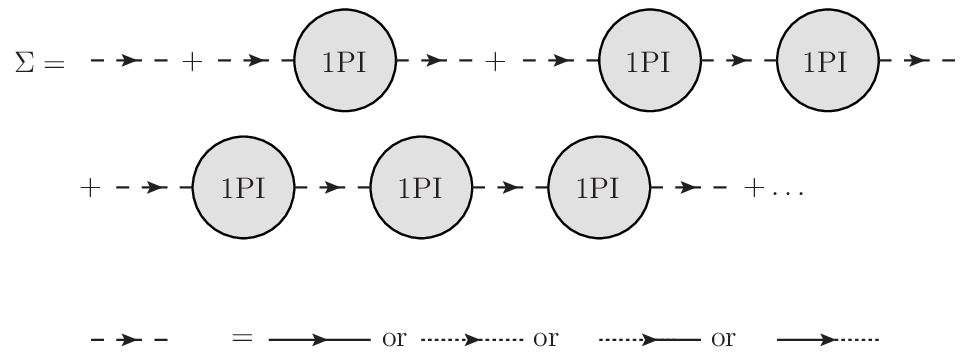}
\caption{The 1PI summation.}
\label{fig:1pisum}
\end{figure}

We can sum up all the 1PI contributions  to infinite orders, as shown in Fig.~\ref{fig:1pisum}. The self-energy is denoted as $\Sigma$, and the 1PI contributions can be written as a matrix,  as in  Eq.~(\ref{eq.3.14}).

The equation in Fig.~\ref{fig:1pisum} can be written as
\begin{equation}
\begin{split}
&\Sigma (\omega_q,q^2) = \sum _{\substack{n=0}}^{\infty}D(\omega_q,{\bf q})\cdot \left(\Pi(\omega_q,q^2) \cdot D(\omega_q,{\bf q})\right)^n=D(\omega_q,{\bf q})\cdot \left(I-\Pi(q^2) \cdot D(\omega_q,{\bf q})\right)^{-1}\\
&=\left(D(\omega_q,{\bf q})^{-1}-\Pi(\omega_q,q^2)\right)^{-1},
\end{split}
\label{eq.3.16}
\end{equation}
where $I$ is the identity matrix, $D(\omega_q, q^2)$ is defined in Eq.~(\ref{eq.2.9}). Eq.~(\ref{eq.3.16}) is the well-known Dyson equation. For simplicity, we only give the result at ${\bf q}=0$, which can be written as
\begin{equation}
\begin{split}
&\Sigma(\omega_q, q^2) \equiv \left(\begin{array}{cc}\Sigma _{11}(\omega_q, q^2)&\Sigma _{21}(\omega_q, q^2)\\ \Sigma_{12}(\omega_q, q^2) & \Sigma_{22}(\omega_q, q^2)\end{array}\right),
\end{split}
\label{eq.3.17}
\end{equation}
with
\begin{equation}
\begin{split}
&\Sigma_{11}(\omega_q, q^2=0)=16 \pi  g^2 v^2 \omega _q \sec ^{-1}\left(\frac{2 g v^2}{\omega _q}\right)/\left[\pi ^2 \left(g^2 v^2+8 \omega _q\right){}^2 \sqrt{4 g^2 v^4-\omega _q^2}\right.\\
&\left.-4 g^3 v^4 \sec ^{-1}\left(\frac{2 g v^2}{\omega _q}\right) \left(g \sqrt{4 g^2 v^4-\omega _q^2} \sec ^{-1}\left(\frac{2 g v^2}{\omega _q}\right)+2 (g+12 \pi ) \omega _q\right)\right],
\end{split}
\label{eq.3.18}
\end{equation}
\begin{equation}
\begin{split}
&\Sigma_{22}(\omega_q, q^2=0)=\frac{16 \pi  g v^2}{\omega _q }\times \left\{\left[2 \sqrt{4 g^2 v^4-\omega _q^2} \left(2 \pi  g^2 v^2-g \omega _q+4 \pi  \omega _q\right)\right.\right.\\
&\left.\left.+g \left(\omega _q^2-20 g^2 v^4\right) \sec ^{-1}\left(\frac{2 g v^2}{\omega _q}\right)\right]/\left[\pi ^2 \left(g^2 v^2+8 \omega _q\right){}^2 \sqrt{4 g^2 v^4-\omega _q^2}\right.\right.\\
&\left.\left.-4 g^3 v^4 \sec ^{-1}\left(\frac{2 g v^2}{\omega _q}\right) \left(g \sqrt{4 g^2 v^4-\omega _q^2} \sec ^{-1}\left(\frac{2 g v^2}{\omega _q}\right)+2 (g+12 \pi ) \omega _q\right)\right]\right\},
\end{split}
\label{eq.3.19}
\end{equation}
\begin{equation}
\begin{split}
&\Sigma_{12}(\omega_q, q^2=0)=-\Sigma_{21}(\omega_q, q^2=0)=-8 \pi\left[\pi  \left(g^2 v^2+8 \omega _q\right) \sqrt{4 g^2 v^4-\omega _q^2}\right.\\
&\left.-8 g^3 v^4 \sec ^{-1}\left(\frac{2 g v^2}{\omega _q}\right)\right]/\left[\pi ^2 \left(g^2 v^2+8 \omega _q\right){}^2 \sqrt{4 g^2 v^4-\omega _q^2}\right.\\
&\left.-4 g^3 v^4 \sec ^{-1}\left(\frac{2 g v^2}{\omega _q}\right) \left(g \sqrt{4 g^2 v^4-\omega _q^2} \sec ^{-1}\left(\frac{2 g v^2}{\omega _q}\right)+2 (g+12 \pi ) \omega _q\right)\right].
\end{split}
\label{eq.3.20}
\end{equation}

The spectrum $\omega (q)$ can be given as the poles of the self-energy~\cite{EFT3},
\begin{equation}
\begin{split}
&\det \left(D(\omega,{\bf q})^{-1}-\Pi(\omega,q)\right)=0.
\end{split}
\label{eq.3.21}
\end{equation}

We find
\begin{equation}
\begin{split}
&\lim _{\omega\to 0}\left[\det \left(D(\omega_q,{\bf q}=0)^{-1}-\Pi(\omega_q,q^2=0)\right)\right]=0.\\
\end{split}
\label{eq.3.22}
\end{equation}
which implies that  $\omega (q^2=0)=0$ is a solution of Eq.~(\ref{eq.3.21}). Therefore there is no gap in  the spectrum of $\psi$, respecting the Hugenholz-Pines theorem~\cite{HugenholzPines}.

The correlation function $\chi _{\psi_1\psi_1}$ can be obtained as
\begin{equation}
\begin{split}
&\chi _{\psi_1\psi_1}(\omega_q, {\bf q})=\Sigma_{11}(\omega_q, q^2=0).
\end{split}
\label{eq.3.23}
\end{equation}

\begin{figure}
\includegraphics[scale=0.7]{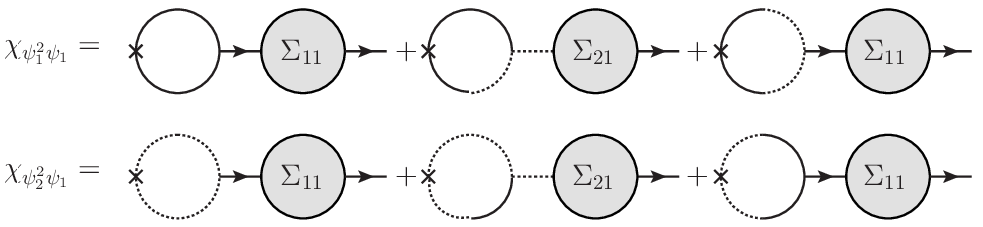}
\caption{The 1PI summation.}
\label{fig:crosssum1}
\end{figure}

\begin{figure}
\includegraphics[scale=0.8]{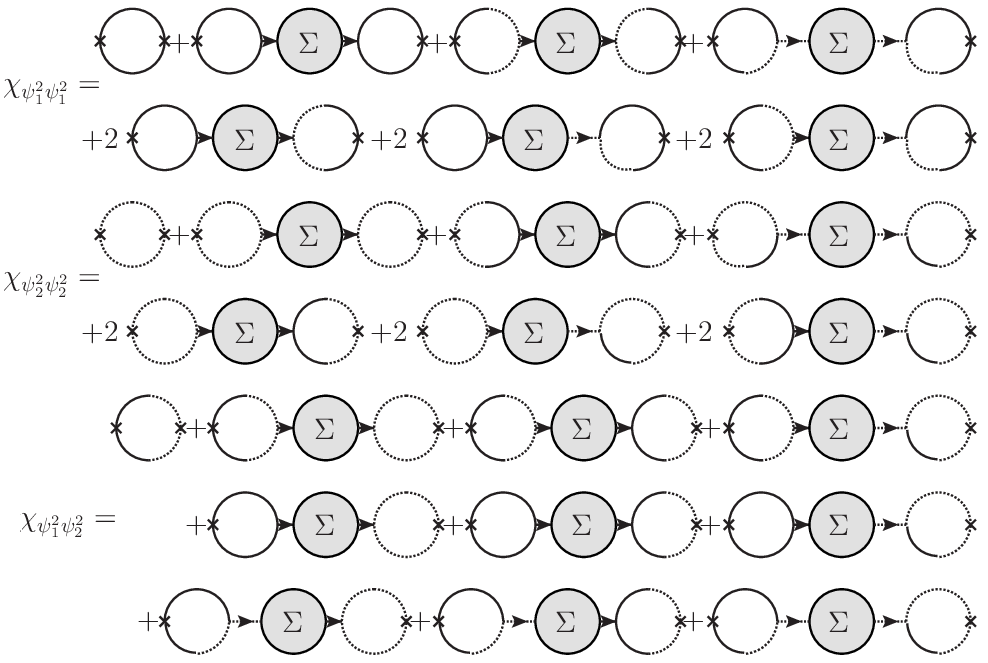}
\caption{The 1PI summation.}
\label{fig:crosssum2}
\end{figure}

The 1PI summation of cross-susceptibilities   are  shown in Figs.~\ref{fig:crosssum1} and \ref{fig:crosssum2}.
The diagrams in Fig.~\ref{fig:crosssum1} represent
\begin{equation}
\begin{split}
&\chi_{\psi_1^2\psi_1}(\omega_q,{\bf q})+\chi_{\psi_2^2\psi_1}(\omega_q,{\bf q})=-2\frac{6gv}{\sqrt{2}}f_a^p(\omega_q,q^2)\Sigma _{11}(\omega_q,q^2)-2\frac{2gv}{\sqrt{2}}f_f^p(\omega_q,q^2)\Sigma _{21}(\omega_q,q^2)\\
&-2\frac{2gv}{\sqrt{2}}f_c^p(\omega_q,q^2)\Sigma _{11}(\omega_q,q^2)-2\frac{2gv}{\sqrt{2}}f_b^p(\omega_q,q^2)\Sigma _{11}(\omega_q,q^2)+2\frac{2gv}{\sqrt{2}}f_g^p(\omega_q,q^2)\Sigma _{21}(\omega_q,q^2)\\
&-2\frac{6gv}{\sqrt{2}}f_c^p(\omega_q,q^2)\Sigma _{11}(\omega_q,q^2).\\
\end{split}
\label{eq.3.24}
\end{equation}

The diagrams in Fig.~\ref{fig:crosssum2} represent
\begin{equation}
\begin{split}
&\chi_{\psi_1^2\psi_1^2}(\omega_q,q^2)=4f_a^p(\omega_q,q^2)+8g^2v^2\left(3f_a^p(\omega_q,q^2)+f_c^p(\omega_q,q^2)\right)^2\Sigma_{11}(\omega_q,q^2)\\
&-8g^2v^2\left(f_f^p(\omega_q,q^2)\right)^2\Sigma _{22}(\omega_q,q^2)-16g^2v^2\left(3f_a^p(\omega_q,q^2)+f_c^p(\omega_q,q^2)\right)f_f^p(\omega_q,q^2)\Sigma _{12}(\omega_q,q^2),\\
&\chi_{\psi_2^2\psi_2^2}(\omega_q,q^2)=4f_b^p(\omega_q,q^2)+8g^2v^2\left(f_b^p(\omega_q,q^2)+3f_c^p(\omega_q,q^2)\right)^2\Sigma_{11}(\omega_q,q^2)\\
&-8g^2v^2\left(f_g^p(\omega_q,q^2)\right)^2\Sigma _{22}(\omega_q,q^2)+16g^2v^2\left(f_b^p(\omega_q,q^2)+3f_c^p(\omega_q,q^2)\right)f_g^p(\omega_q,q^2)\Sigma _{12},\\
&\chi_{\psi_1^2\psi_2^2}(\omega_q,q^2)=4f_c^p(\omega_q,q^2)+8g^2v^2\left(f_c^p(\omega_q,q^2)+3f_a^p(\omega_q,q^2)\right)\times \left[\Sigma_{11}(\omega_q,q^2)f_b^p(\omega_q,q^2)\right.\\
&\left.+3\Sigma _{11}(\omega_q,q^2)f_c^p(\omega_q,q^2)+\Sigma_{12}f_g^p(\omega_q,q^2)\right]+8g^2v^2f_f^p(\omega_q,q^2)\times \left[\Sigma_{21}(\omega_q,q^2)f_b^p(\omega_q,q^2)\right.\\
&\left.+3\Sigma _{21}(\omega_q,q^2)f_c^p(\omega_q,q^2)+\Sigma_{22}(\omega_q,q^2)f_g^p(\omega_q,q^2)\right].
\end{split}
\label{eq.3.25}
\end{equation}

Using Eqs.~(\ref{eq.2.13}), (\ref{eq.3.23}), (\ref{eq.3.24}) and (\ref{eq.3.25}), we find
\begin{equation}
\begin{split}
&\chi _{\rho\rho}(\omega_q,{\bf q}=0)=\left[64 \pi  g^2 v^4 \omega _q \sec ^{-1}\left(\frac{2 g v^2}{\omega _q}\right)\right]/\left[64 \pi ^2 \omega _q^2 \sqrt{4 g^2 v^4-\omega _q^2}\right.\\
&\left.-8 g^2 v^2 \omega _q \left(g (g+12 \pi ) v^2 \sec ^{-1}\left(\frac{2 g v^2}{\omega _q}\right)-2 \pi ^2 \sqrt{4 g^2 v^4-\omega _q^2}\right)\right.\\
&\left.+g^4 v^4 \sqrt{4 g^2 v^4-\omega _q^2} \left(\pi ^2-4 \sec ^{-1}\left(\frac{2 g v^2}{\omega _q}\right){}^2\right)\right].
\end{split}
\label{eq.3.26}
\end{equation}

In experiments, the spectral function is normalized after being measured~\cite{2DOpticalLattice}. We find that  after normalization, the spectral functions $\chi ''_{\psi _1\psi _1}(\omega_q,{\bf q}=0)$ and $\chi ''_{\rho\rho}(\omega_q,{\bf q}=0)$ are the same as each other. In the rest of the paper, we only concentrate on $\chi ''_{\psi _1\psi _1}(\omega_q,{\bf q}=0)$.

\section{\label{sec:4}Numerical results}

To obtain the numerical results, we need to match the coupling constant $g$. One can match the coupling constant $g$ at tree level and  the leading order of $q^2$, the result is  $g=8\pi a_s$ \cite{EFT3}, where $a_s$ is the s-wave scattering length. However, in experiments, the system is tunable via $j\equiv J/U$, where $J$ is the hopping constant, $U$ is the interaction strength. To obtain the dependence of the parameters on the the hopping constant $J$, we introduce an action  derived from the Bose-Hubbard model by using Hubbard-Stratanovich transformation~\cite{HBM1,HBM2,HBM3},
\begin{equation}
\begin{split}
&S[\psi ^*,\psi]=\int _0^{\beta} d\tau \int d^D x \left\{K_1\psi ^*\frac{\partial}{\partial \tau}\psi+K_2|\frac{\partial}{\partial \tau}\psi |^2 +K_3|\nabla \psi|^2+r|\psi|^2+\frac{u}{2}|\psi|^4+\mathcal{O}(\psi ^6)\right\},
\end{split}
\label{eq.4.1}
\end{equation}
where
\begin{equation}
\begin{split}
&r=\frac{1}{Za^d}\left(\frac{1}{J}-\left(\frac{n_0+1}{n_0U-\mu}+\frac{n_0}{\mu -(n_0-1)U}\right)\right),\\
&n_0=\left\{\begin{array}{cc}0,&\mu / U < 0;\\1,&0<\mu/U<1;\\2,&1<\mu/U<2;\\ \ldots &\end{array}\right.\\
&K_1=-\frac{r}{\mu},\\
\end{split}
\label{eq.4.2}
\end{equation}
where $Z$ is the coordinate number, $a$ is the lattice spacing, $\mu$ is the chemical potential.   Comparing  Eq.~(\ref{eq.4.1}) with Eq.~(\ref{eq.2.1}), one can find that when $K_2=0$,  Eq.~(\ref{eq.4.1}) reduces to Eq.~(\ref{eq.2.1}), which is the EFT.

\begin{figure}
\includegraphics[scale=0.6]{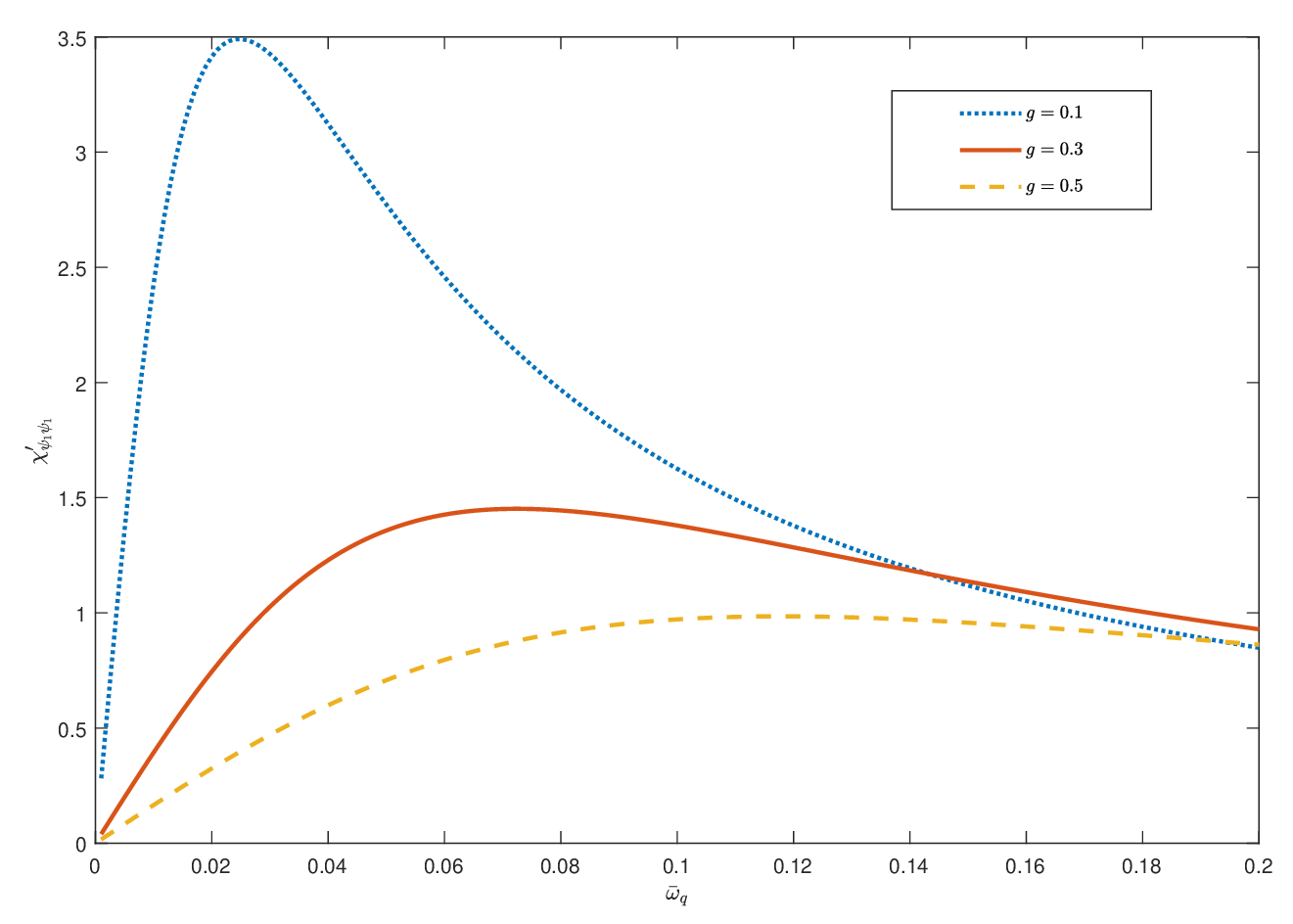}
\caption{The normalized spectral function $\chi ''_{\psi_1\psi_1}(\omega _q{\bf q}=0)$ at $\bar{r}=2$. The dotted line is for $g=0.1$, the solid line is for $g=0.3$ and the dashed line is for $g=0.5$. One can see that the peaks of the spectral functions are broadened continuums rather than sharp peaks. However, similar to $O(2)$ model, when $g$ decreases, the peak  becomes sharper.}
\label{fig:singlechipsiplot}
\end{figure}

Based on the comparison  of Eqs.~(\ref{eq.4.1}) and (\ref{eq.4.2}) with Eq.~(\ref{eq.2.1}), we assume
\begin{equation}
\begin{split}
&r=\alpha \left(\frac{j}{j_c}-1\right),\;\;\bar{r}\equiv \frac{r}{\alpha}=\left(\frac{j}{j_c}-1\right).
\end{split}
\label{eq.4.3}
\end{equation}
where  $\alpha$ is an arbitrary constant parameter with  dimension of $m^2$. Then we can use the  dimensionless variable
\begin{equation}
\begin{split}
&\bar{\omega }\equiv \frac{\omega _q}{\alpha}.
\end{split}
\label{eq.4.4}
\end{equation}
After variable substitution and normalization, $\chi ''_{\psi_1\psi_1}$ depends only  on the  massless parameters $\bar{r}$, $\bar{\omega_q}$ and $g$. The perturbation works only when $g\ll 1$, so we choose $g<1$. The normalized spectral function $\chi ''_{\psi _1\psi _1}(\omega_q,{\bf q}=0)$  is  shown in Fig.~\ref{fig:singlechipsiplot} with the parameter values  $\bar{r}=2$ and $g=0.1$, $g=0.3$ and $g=0.5$.

We find that the peaks of the spectral functions form  broadened  continuums rather than sharp peaks, in consistency with   the experiment~\cite{2DOpticalLattice}. We also find that  similar to  $O(2)$ model, when $g$ decreases, the peak  becomes sharper. This  cannot explain the disappearance of the Higgs mode observed in the experiment~\cite{2DOpticalLattice}.

When $\bar{r}\gg \bar{\omega_q}$, the spectral function can be simplified as
\begin{equation}
\begin{split}
&\chi ''_{\psi_1\psi_1}(\bar{\omega}_q,{\bf q}=0)\approx \frac{1}{N}\frac{4 \pi  g \left(\pi  \bar{r}\left(g^2 \bar{\omega}_q+24 \pi  g \bar{\omega}_q+64 \pi ^2 \bar{\omega}_q\right)-8 \bar{\omega}_q \left(\pi ^2 g \bar{r}\right)\right)}{\bar{r} \left(\left(g^2 \bar{\omega}_q+24 \pi  g \bar{\omega}_q+64 \pi ^2 \bar{\omega}_q\right)^2+\left(8 \pi ^2 g \bar{r}\right)^2\right)}
\end{split}
\label{eq.4.5}
\end{equation}
where $N$ is the normalization factor. So we can find the maximum is at
\begin{equation}
\begin{split}
&\bar{\omega}_q=\frac{8 g \bar{r}\pi ^2}{g^2+24g \pi+64\pi^2}\approx \frac{g}{8}\left(\frac{j}{j_c}-1\right)\\
\end{split}
\label{eq.4.6}
\end{equation}
when $\bar{r}\gg \bar{\omega_q}$ and $g\ll 1$.

\begin{figure}
\includegraphics[scale=0.7]{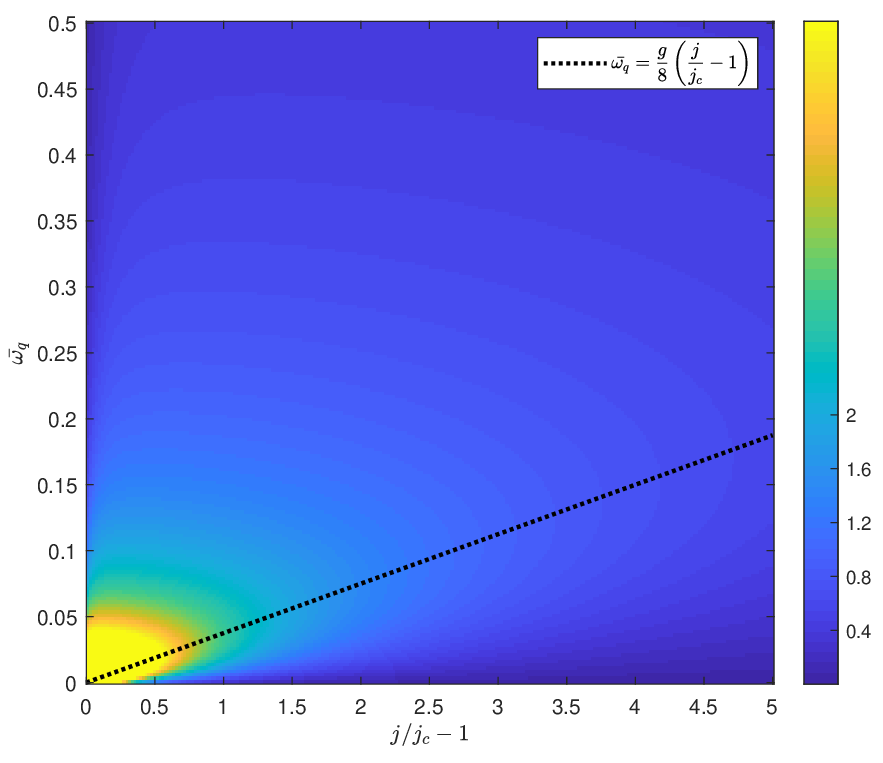}
\caption{The spectral function of longitudinal susceptibility $\chi ''_{\psi_1\psi_1}$ at $g=0.3$. The dashed line is the approximate position of the maximum of the spectral function shown in Eq.~(\ref{eq.4.6}). }
\label{fig:chipsiplot}
\end{figure}

In the experiment, with the increase of  the lattice potential depth, $g$ increases approximately linearly and $J$ decreases exponentially~\cite{Bloch}. The spectral function as a function of $j/j_c$,  while $g$ is kept constant, is shown in  Fig.~\ref{fig:chipsiplot}. One canclearly   see the peak and the energy gap,  as well as the disappearance of the Higgs mode. We find that the spectral function shown in Fig.~\ref{fig:chipsiplot} fits well  the observation in the experiment~\cite{2DOpticalLattice}.

\section{\label{sec:5}Conclusion}

The Higgs  mode discovered in the 2D optical lattice ended the debate  whether the Higgs  mode  can be observed in the 2D neutral superfluid. However, the feature that the peek is a soft continuum above the gap energy rather than a sharp peak, and the disappearance of the response in the ordered phase, cannot be explained using the $O(2)$ model.

In this paper, we have investigated the spectral function of the Higgs  mode by using an EFT model. We calculate the spectral functions of both longitudinal susceptibility $\chi ''_{\psi_1\psi _1}$ and scalar susceptibility $\chi ''_{\rho\rho}$. The spectral functions are obtained as shown in Eqs.~(\ref{eq.3.23}) and (\ref{eq.3.26}), and are drawn in Fig.~\ref{fig:singlechipsiplot} and Fig.~\ref{fig:chipsiplot}.

We find that  the visibility of the Higgs  mode is the same in  longitudinal  and scalar susceptibilities. Our EFT calculation reproduces various experimental features, including   the softness of  the peak of the spectral function and  the disappearance of the response with the increase of  $j/j_c$.

This work is supported by National Natural Science Foundation of China (Grant No. 12075059).

\appendix

\section{\label{sec:a}The results of Feynman diagrams}

\subsection{\label{sec:a.1}Results of some integrals}

Similar to Ref.~\cite{EFT3}, we also use the definition
\begin{equation}
\begin{split}
&I_{m,n}(A^2)\equiv M^{\epsilon}\int \frac{d^Dk}{(2\pi)^D}\frac{k^{2m}}{k^n(k^2+A ^2)^{\frac{n}{2}}}=\frac{M^{\epsilon}A^{D+2m-2n}}{(4\pi)^{\frac{D}{2}}}\frac{\Gamma (\frac{D-n}{2}+m)\Gamma (n-m-\frac{D}{2})}{\Gamma (\frac{D}{2})\Gamma (\frac{n}{2})}.\\
\end{split}
\label{eq.a.1}
\end{equation}

Another integral we need can be defined as
\begin{equation}
\begin{split}
&J_{a,b,c}(A^2,B^2)\equiv M^{\epsilon}\int \frac{d^D k}{(2\pi)^D}\frac{1}{(k^2)^a(k^2+A^2)^b(4k^2(k^2+A^2)+B^2)^c}.\\
\end{split}
\label{eq.a.2}
\end{equation}

It can be calculated in Mellin-Barnes representation~\cite{MellinBarnes}, as
\begin{equation}
\begin{split}
&J_{a,b,c}=\frac{M^{\epsilon}}{2^{2c}2\pi i}\int _{-i\infty}^{i\infty}dz\frac{\Gamma (c+z)\Gamma (-z)}{\Gamma (c)}\int \frac{d^D k}{(2\pi)^D}\frac{\left(\frac{B^2}{4}\right)^z}{(k^2)^a(k^2+A^2)^b(k^2(k^2+A^2))^{c+z}}.\\
\end{split}
\label{eq.a.3}
\end{equation}

With the help of $I_{m,n}$ calculated in Eq.~(\ref{eq.a.1}), it can be written as
\begin{equation}
\begin{split}
&J_{a,b,c}(A^2,B^2)=\frac{M^{\epsilon}\left(\frac{A^2}{2}\right)^{\frac{D}{2}-a-b-2c}}{2^{2c+1}\Gamma (c)\Gamma (\frac{D}{2})(4\pi)^{\frac{D}{2}}\sqrt{\pi}}\frac{1}{2\pi i}\int _{-i\infty}^{i\infty}dz\left(\frac{B^2}{A^4}\right)^z\\
&\times \frac{\Gamma (c+z)\Gamma (-z)\Gamma (\frac{D}{2}-a-c-z)\Gamma (\frac{a+b}{2}+c-\frac{D}{4}+z)\Gamma (\frac{a+b+1}{2}+c-\frac{D}{4}+z)}{\Gamma (b+c+z)}.\\
\end{split}
\label{eq.a.4}
\end{equation}

For convenience, we define
\begin{equation}
\begin{split}
&j(a,b,c,d,e)\equiv\frac{1}{2\pi i}\int _{-i\infty}^{i\infty}dz \frac{\Gamma(a+z)\Gamma (b+z)\Gamma (c+z)\Gamma (d-z)\Gamma (-z)}{\Gamma (e+z)}t^z,\\
\end{split}
\label{eq.a.5}
\end{equation}
which is calculated by closing the contour of the integral and using
\begin{equation}
\begin{split}
&{\rm Res} (\Gamma (a\pm n),z=\mp (n+a))=\pm \frac{(-1)^n}{n!}.\\
\end{split}
\label{eq.a.6}
\end{equation}
It is obtained that
\begin{equation}
\begin{split}
&j(a,b,c,d,e)=\sum _{n=0}^{\infty}\left(\Gamma(d)\Gamma(1-d)\frac{\Gamma(a+n)\Gamma (b+n)\Gamma (c+n)}{\Gamma (e+n)\Gamma (1-d+n)}\frac{t^n}{n!}\right.\\
&\left.+t^d\Gamma(-d)\Gamma(1+d)\frac{\Gamma(a+d+n)\Gamma (b+d+n)\Gamma (c+d+n)}{\Gamma (e+d+n)\Gamma (1+d+n)}\frac{t^n}{n!}\right),\\
\end{split}
\label{eq.a.7}
\end{equation}
where we have used  the relation
\begin{equation}
\begin{split}
&\Gamma (x-n)=(-1)^n \frac{\Gamma (x)\Gamma (1-x)}{\Gamma (1-x+n)},
\end{split}
\label{eq.a.8}
\end{equation}
when $n$ is an integer.

Then using the definition of Hypergeometric function
\begin{equation}
\begin{split}
&\;_pF_q\left(\left.\begin{array}{c}a_1,a_2,...,a_p\\b_1,b_2,...,b_q\end{array}\right|x\right)=\sum _{n=0}^{\infty} \frac{\prod_{i=1}^{p} (a_i)_n}{\prod _{j=1}^{q} (b_j)_n}\frac{x^n}{n!},
\end{split}
\label{eq.a.9}
\end{equation}
we find
\begin{equation}
\begin{split}
&j(a,b,c,d,e)=\frac{\Gamma(a)\Gamma(b)\Gamma(c)\Gamma(d)}{\Gamma(e)}\;_3F_2\left(\left.\begin{array}{c}a,b,c\\ e,1-d\end{array}\right|t\right)\\
&+t^d\frac{\Gamma(a+d)\Gamma(b+d)\Gamma(c+d)\Gamma(-d)}{\Gamma(e+d)}\;_3F_2\left(\left.\begin{array}{c}a+d,b+d,c+d\\ e+d,1+d\end{array}\right|t\right).\\
\end{split}
\label{eq.a.10}
\end{equation}

With the help of $j(a,b,c,d,e)$ calculated in Eq.~(\ref{eq.a.10}), $J_{a,b,c}$ can be written as
\begin{equation}
\begin{split}
&J_{a,b,c}=\frac{M^{\epsilon}\left(A^2\right)^{\frac{D}{2}-a-b-2c}}{2^{2c}\Gamma (\frac{D}{2})(4\pi)^{\frac{D}{2}}}\\
&\times \left(\frac{\Gamma(a+b+2c-\frac{D}{2})\Gamma(\frac{D}{2}-a-c)}{\Gamma(b+c)}\;_3F_2\left(\left.\begin{array}{c}c,\frac{a+b}{2}+c-\frac{D}{4},\frac{a+b+1}{2}+c-\frac{D}{4}\\b+c,1+a+c-\frac{D}{2}\end{array}\right|\frac{B^2}{A^4}\right)\right.\\
&+\left(\frac{B^2}{4A^4}\right)^{\frac{D}{2}-a-c}\left.\frac{\Gamma(\frac{D}{2}-a)\Gamma(a+c-\frac{D}{2})}{\Gamma (c)}\;_3F_2\left(\left.\begin{array}{c}\frac{D}{2}-a,\frac{b-a}{2}+\frac{D}{4},\frac{b-a+1}{2}+\frac{D}{4}\\\frac{D}{2}+b-a,1-a-c+\frac{D}{2}\end{array}\right|\frac{B^2}{A^4}\right)\right).\\
\end{split}
\label{eq.a.11}
\end{equation}

When using DR to regulate the UV divergences we need to calculate the $\epsilon$-expansion of the Hypergeometric function which can be written as
\begin{equation}
\begin{split}
&h\equiv \frac{A^{\epsilon}\Gamma (x_1+\alpha _1 \epsilon)\Gamma (\alpha _2 \epsilon)}{\Gamma (x_2+\alpha _3 \epsilon)\times (x_3+\alpha _4 \epsilon)}\;_3F_2\left(\left.\begin{array}{c}y_1,y_2+\beta _1 \epsilon, \beta_2 \epsilon\\ y_3,y_4+\beta_3\epsilon\end{array}\right|t\right).
\end{split}
\label{eq.a.12}
\end{equation}
Using the definition Eq.~(\ref{eq.a.7}), we find
\begin{equation}
\begin{split}
&h=\frac{A^{\epsilon}\Gamma (x_1+\alpha _1 \epsilon)\Gamma (\alpha _2 \epsilon)}{\Gamma (x_2+\alpha _3 \epsilon)\times (x_3+\alpha _4 \epsilon)}+\frac{A^{\epsilon}\Gamma (x_1+\alpha _1 \epsilon)\Gamma (\alpha _2 \epsilon)}{\Gamma (x_2+\alpha _3 \epsilon)\times (x_3+\alpha _4 \epsilon)}\sum _{n=1}^{\infty} \frac{\frac{\Gamma(y_1+n)}{\Gamma (y_1)}\frac{\Gamma(y_2+\beta _1 \epsilon+n)}{\Gamma (y_2+\beta _1 \epsilon)}\frac{\Gamma(\beta _2\epsilon+n)}{\Gamma (\beta _2 \epsilon)}}{\frac{\Gamma(y_3+n)}{\Gamma (y_3)}\frac{\Gamma(y_4+\beta _3 \epsilon+n)}{\Gamma (y_4+\beta _3 \epsilon)}}\frac{t^n}{n!}.\\
\end{split}
\label{eq.a.13}
\end{equation}
Then we can expand the Gamma function around $\epsilon \to 0$ in each term and gather the summation, and obtain
\begin{equation}
\begin{split}
&h=\frac{1}{\epsilon}\frac{\Gamma (x_1)}{\alpha _2x_3\Gamma (x_2)}+\frac{\Gamma (x_1)\left(\log(A)+\alpha _1 \psi ^{(0)}(x_1)-\alpha _3 \psi ^{(0)}(x_2)-\gamma _E \alpha _2\right)}{\alpha _2x_3\Gamma (x_2)}-\frac{\alpha _4 \Gamma (x_1)}{\alpha _2 x_3^2\Gamma (x_2)}\\
&+t\frac{\beta _2\Gamma (x_1)}{x_3\alpha _2\Gamma (x_2)}\sum _{n=0}^{\infty}\Gamma (n+1) \frac{\frac{\Gamma(y_1+1+n)}{\Gamma (y_1)}\frac{\Gamma(y_2+1+n)}{\Gamma (y_2)}}{\frac{\Gamma(y_3+1+n)}{\Gamma (y_3)}\frac{\Gamma(y_4+1+n)}{\Gamma (y_4)}}\frac{t^n}{\Gamma (n+2)}+\mathcal{O}(\epsilon)\\
&=\frac{1}{\epsilon}\frac{\Gamma (x_1)}{\alpha _2x_3\Gamma (x_2)}+\frac{\Gamma (x_1)\left(\log(A)+\alpha _1 \psi ^{(0)}(x_1)-\alpha _3 \psi ^{(0)}(x_2)-\gamma _E \alpha _2\right)}{\alpha _2x_3\Gamma (x_2)}-\frac{\alpha _4 \Gamma (x_1)}{\alpha _2 x_3^2\Gamma (x_2)}\\
&+t\frac{\beta _2\Gamma (x_1)}{x_3\alpha _2\Gamma (x_2)}\frac{\Gamma (y_3)\Gamma (y_4)\Gamma (y_1+1)\Gamma (y_2+1)}{\Gamma (y_1)\Gamma (y_2)\Gamma (y_3+1)\Gamma (y_4+1)}\;_4F_3\left(\left.\begin{array}{c}1,1,y_1+1,y_2+1\\2,y_3+1,y_4+1\end{array}\right|t\right)+\mathcal{O}(\epsilon).\\
\end{split}
\label{eq.a.14}
\end{equation}
where $\gamma _E$ is the Eular constant, $\psi ^{(0)}(x)$ is the digamma function.

\subsection{\label{sec:a.2}Tadpole diagrams}

\begin{figure}
\includegraphics[scale=0.7]{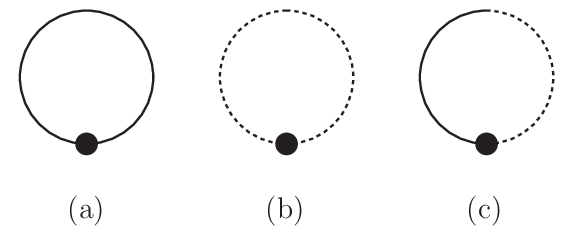}
\caption{All tadpole diagrams at 1-loop level.}
\label{fig:alltadpoles}
\end{figure}

All the tadpole diagrams at 1-loop level are drawn in Fig.~\ref{fig:alltadpoles}. The diagrams in Fig.~\ref{fig:alltadpoles}.~(a), (b) and (c) are denoted as $f_a^t$, $f_b^t$ and $f_c^t$, and  can be written as
\begin{equation}
\begin{split}
&f_a^t=\frac{1}{2}M^{\epsilon}\int \frac{d\omega}{2\pi}\int \frac{d^Dk}{(2\pi)^D}\frac{k^2}{\omega ^2+\epsilon ^2(k)},\\
&f_b^t=\frac{1}{2}M^{\epsilon}\int \frac{d\omega}{2\pi}\int \frac{d^Dk}{(2\pi)^D}\frac{k^2+2gv^2}{\omega ^2+\epsilon ^2(k)},\\
&f_c^t=M^{\epsilon}\int \frac{d\omega}{2\pi}\int \frac{d^D k}{(2\pi)^D}\frac{\omega}{\omega ^2 +\epsilon ^2(k)}.\\
\end{split}
\label{eq.a.15}
\end{equation}

We first integrate over $\omega$, then express the result in terms of  $I_{m,n}$ defined in Eq.~(\ref{eq.a.1}),
\begin{equation}
\begin{split}
&f_a^t=\frac{1}{4}I_{1,1}(2gv^2),\;\;\;f_b^t=\frac{1}{4}I_{-1,-1}(2gv^2),\;\;\;f_c^t=0.\\
\end{split}
\label{eq.a.16}
\end{equation}

In $D=2-\epsilon$ dimensions, using Eq.~(\ref{eq.a.1}), we find
\begin{equation}
\begin{split}
&f_a^t=\frac{2gv^2}{4}\left(-\frac{N_{\rm UV}}{8\pi}+\frac{1}{8\pi}\right),\;\;\;f_b^t=\frac{2gv^2}{4}\left(\frac{N_{\rm UV}}{8\pi}+\frac{1}{8\pi}\right).\\
\end{split}
\label{eq.a.17}
\end{equation}

\subsection{\label{sec:a.3}Polarization diagrams}

\begin{figure}
\includegraphics[scale=0.7]{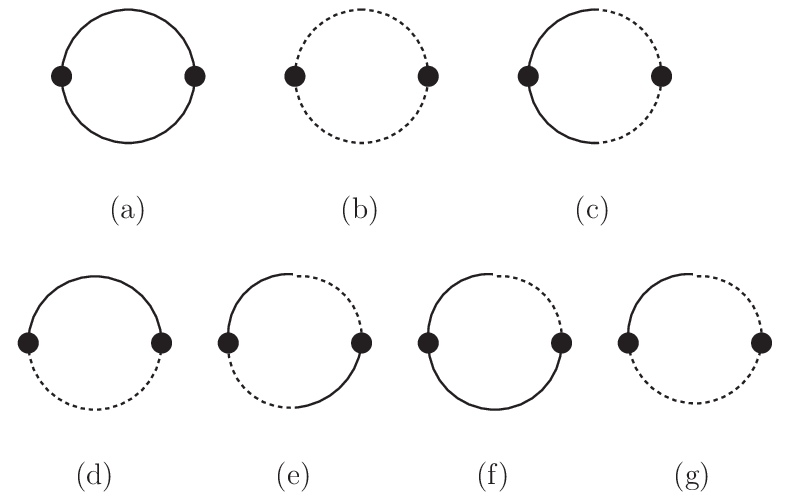}
\caption{Other diagrams at 1-loop level.}
\label{fig:allpolarization}
\end{figure}

Other 1-loop diagrams we need are listed in Fig.~\ref{fig:allpolarization}, and the diagrams in Fig.~\ref{fig:allpolarization}.~(a), (b), (c), (d), (e), (f) and (c) are denoted as $f_a^p(q^2)$, $f_b^p(q^2)$, $f_c^p(q^2)$, $f_d^p(q^2)$, $f_e^p(q^2)$, $f_f^p(q^2)$ and $f_g^p(q^2)$, and can be written as
\begin{equation}
\begin{split}
&f_a^p(\omega _q,q^2)=\frac{1}{2}M^{\epsilon}\int \frac{d\omega}{2\pi}\int \frac{d^Dk}{(2\pi)^D}\frac{k^2}{\omega ^2+\epsilon ^2(k)}\frac{(k+q)^2}{(\omega+\omega_q) ^2+\epsilon ^2(k+q)},\\
&f_b^p(\omega _q,q^2)=\frac{1}{2}M^{\epsilon}\int \frac{d\omega}{2\pi}\int \frac{d^Dk}{(2\pi)^D}\frac{k^2+2gv^2}{\omega ^2+\epsilon ^2(k)}\frac{(k+q)^2+2gv^2}{(\omega+\omega_q) ^2+\epsilon ^2(k+q)},\\
&f_c^p(\omega _q,q^2)=-\frac{1}{2}M^{\epsilon}\int \frac{d\omega}{2\pi}\int \frac{d^Dk}{(2\pi)^D}\frac{\omega (\omega+\omega_q)}{(\omega ^2+\epsilon ^2(k))((\omega+\omega_q) ^2+\epsilon ^2(k+q))},\\
&f_d^p(\omega _q,q^2)=M^{\epsilon}\int \frac{d\omega}{2\pi}\int \frac{d^Dk}{(2\pi)^D}\frac{k^2}{\omega ^2+\epsilon ^2(k)}\frac{(k+q)^2+2gv^2}{(\omega+\omega_q) ^2+\epsilon ^2(k+q)},\\
&f_e^p(\omega _q,q^2)=M^{\epsilon}\int \frac{d\omega}{2\pi}\int \frac{d^Dk}{(2\pi)^D}\frac{\omega (\omega+\omega_q)}{(\omega ^2+\epsilon ^2(k))((\omega+\omega_q) ^2+\epsilon ^2(k+q))}=-2f_c^p(q^2),\\
&f_f^p(\omega _q,q^2)=M^{\epsilon}\int \frac{d\omega}{2\pi}\int \frac{d^Dk}{(2\pi)^D}\frac{k^2}{\omega ^2+\epsilon ^2(k)}\frac{-(\omega+\omega_q)}{(\omega+\omega_q) ^2 +\epsilon ^2(k+q)},\\
&f_g^p(\omega _q,q^2)=M^{\epsilon}\int \frac{d\omega}{2\pi}\int \frac{d^Dk}{(2\pi)^D}\frac{k^2+2gv^2}{\omega ^2+\epsilon ^2(k)}\frac{-(\omega+\omega_q)}{(\omega+\omega_q) ^2 +\epsilon ^2(k+q)}.\\
\end{split}
\label{eq.a.18}
\end{equation}

We also calculate those integrals at long wave length limit as  in Ref.~\cite{EFT3}, that is,  after integrating over $\omega$, we expand the result at $q^2\to 0$ before integrating over $k$.

Take $f_c^p(q^2)$ as an example, after Feynman parameter, $f_c^p(q^2)$ can be written as
\begin{equation}
\begin{split}
&f_c^p(\omega _q, q^2)=-\frac{1}{2}\int _0^1dx\int \frac{d\omega}{2\pi}\int \frac{d^Dk}{(2\pi)^D}\left(\frac{\omega^2+(1-2x)\omega\omega_q-x(1-x)\omega _q^2}{\left(\omega ^2+x\epsilon ^2(k+q)+(1-x)\epsilon ^2(k)+x(1-x)\omega_q^2\right)^2}\right).\\
\end{split}
\label{eq.a.19}
\end{equation}

The terms with odd powers of $\omega$ do not contribute. Hence the integral can be written as
\begin{equation}
\begin{split}
&f_c^p(\omega _q,q^2)=f_{c_1}^p+f_{c_2}^p,\\
&f_{c_1}^p=-\frac{1}{2}\int _0^1dx\int \frac{d\omega}{2\pi}\int \frac{d^Dk}{(2\pi)^D}\left(\frac{\omega ^2}{\left(\omega ^2+x\epsilon ^2(k+q)+(1-x)\epsilon ^2(k)+x(1-x)\omega _q^2\right)^2}\right),\\
&I_{c_2}^p=-\frac{1}{2}\int _0^1dx\int \frac{d\omega}{2\pi}\int \frac{d^Dk}{(2\pi)^D}\left(\frac{-x(1-x)\omega _q^2}{\left(\omega ^2+x\epsilon ^2(k+q)+(1-x)\epsilon ^2(k)+x(1-x)\omega_q^2\right)^2}\right).\\
\end{split}
\label{eq.a.20}
\end{equation}
After integrating over $\omega$, $f_{c_1}^p$ can be written as
\begin{equation}
\begin{split}
&f_{c_1}^p=-\frac{1}{8}\int _0^1dx\int \frac{d^Dk}{(2\pi)^D}\frac{1}{(x\epsilon ^2(k+q)+(1-x)\epsilon ^2(k)+x(1-x)\omega_q^2)^{\frac{1}{2}}}.\\
\end{split}
\label{eq.a.21}
\end{equation}
By using integration-by-part~(IBP) recursive relation~\cite{IBP},
\begin{equation}
\begin{split}
&D\int d^D k f(k)+\int d^D k \left(k\cdot \frac{\partial }{\partial k}f(k)\right)=0, \\
\end{split}
\label{eq.a.22}
\end{equation}
we obtain
\begin{equation}
\begin{split}
&f_{c_1}^p=-\frac{1}{8D}\int _0^1dx\int \frac{d^Dk}{(2\pi)^D}\frac{1}{(x\epsilon ^2(k+q)+(1-x)\epsilon ^2(k)+x(1-x)\omega _q^2)^{\frac{3}{2}}}\\
&\times \left((1-x)\times \left(k^4+k^2(k^2+2gv^2)\right)+x\times \left(k\cdot (k+q)\left(2(k+q)^2+2gv^2\right)\right)\right).\\
\end{split}
\label{eq.a.23}
\end{equation}
Then by integrating over $x$,  we write the result  as
\begin{equation}
\begin{split}
&f_{c_1}^p=-\frac{1}{4D}\int \frac{d^Dk}{(2\pi)^D}\left(\frac{k^4+k^2(k^2+2gv^2)}{\epsilon (k)\left((\epsilon(k)+\epsilon(k+q))^2+\omega_q^2\right)}\right.\\
&\left.+\frac{k\cdot (k+q)\left(2(k+q)^2+2gv^2\right)}{\epsilon (k+q)\left((\epsilon(k)+\epsilon(k+q))^2+\omega_q^2\right)}\right).\\
\end{split}
\label{eq.a.24}
\end{equation}
Then we use the long wavelength approximation, and expand $f_{c_1}^p$ around $q^2\to 0$, and obtain
\begin{equation}
\begin{split}
&f_{c_1}^p=-\frac{1}{4D}\left\{2\left[J_{-\frac{3}{2},\frac{1}{2},1}+J_{-\frac{1}{2},-\frac{1}{2},1}\right]
+\frac{q^2}{2D}\left[(1-D)\omega_q^4m^6J_{\frac{1}{2},\frac{5}{2},3}\right.\right.\\
&\left.\left.-\omega_q^2m^4\left((4-D)\omega_q^2-4(4D-3)m^4\right)J_{-\frac{1}{2},\frac{5}{2},3}\right.\right.\\
&\left.\left.+4\left((3-16D)\omega_q^2m^6+12(2-D)m^{10}\right)J_{-\frac{3}{2},\frac{5}{2},3}\right.\right.\\
&\left.\left.+16\left(-7(1+D)\omega_q^2m^4+(32-17D)m^8\right)J_{-\frac{5}{2},\frac{5}{2},3}\right.\right.\\
&\left.\left.+16\left(-(11+16D)\omega_q^2m^2+(66-41D)m^6\right)J_{-\frac{7}{2},\frac{5}{2},3}\right.\right.\\
&\left.\left.+16\left(-2(2+D)\omega_q^2+(76-51D)m^46\right)J_{-\frac{9}{2},\frac{5}{2},3}\right.\right.\\
&\left.\left.+64(13-8D)m^2J_{-\frac{11}{2},\frac{5}{2},3}+128(D-2)
J_{-\frac{13}{2},\frac{5}{2},3}\right]\right\}+\mathcal{O}(q^4).\\
\end{split}
\label{eq.a.25}
\end{equation}
In above, we have used the relation
\begin{equation}
\begin{split}
&\int d^Dk (k\cdot q)^2=\int d^Dk \frac{k^2q^2}{D},
\end{split}
\label{eq.a.26}
\end{equation}
and  defined $m^2\equiv 2gv^2$ for convenience.

Using the same procedure as for $f_{c_2}^p$, we find
\begin{equation}
\begin{split}
&f_{c_2}^p=\frac{\omega_q^2}{2D}\left\{2\left[J_{-\frac{3}{2},\frac{1}{2},2}+J_{-\frac{1}{2},-\frac{1}{2},2}\right]-\frac{q^2}{2D}\left[(D-1)\omega_q^4m^6J_{\frac{1}{2},\frac{5}{2},4}\right.\right.\\
&\left.\left.-\omega_q^2m^4\left((4-D)\omega_q^2-8(2-3D)m^4\right)J_{-\frac{1}{2},\frac{5}{2},4}\right.\right.\\
&\left.\left.-16m^6\left(-(7D+1)\omega_q^2+5(D-3)m^4\right)J_{-\frac{3}{2},\frac{5}{2},4}\right.\right.\\
&\left.\left.-8\left(-(32+27D)\omega_q^2m^4+2(90-31D)m^8\right)J_{-\frac{5}{2},\frac{5}{2},4}\right.\right.\\
&\left.\left.-16\left(-2(11+6D)\omega_q^2m^2+(223-79D)m^6\right)J_{-\frac{7}{2},\frac{5}{2},4}\right.\right.\\
&\left.\left.-16\left(-4(2+D)\omega_q^2+(300-101D)m^4\right)J_{-\frac{9}{2},\frac{5}{2},4}\right.\right.\\
&\left.\left.+128(8D-27)m^2J_{-\frac{11}{2},\frac{5}{2},4}+256(D-4)J_{-\frac{13}{2},\frac{5}{2},4}\right]\right\}+\mathcal{O}(q^4).\\
\end{split}
\label{eq.a.27}
\end{equation}

In $D=2-\epsilon$ dimensions, using Eqs.~(\ref{eq.a.11}) and (\ref{eq.a.11}), we find
\begin{equation}
\begin{split}
&f_c^p(\omega _q,q^2)=-\frac{1}{4}\left\{\frac{N_{\rm UV}}{8\pi}-\frac{\omega _q \cos ^{-1}\left(\frac{\omega _q}{m^2}\right)}{8 \pi  \sqrt{m^4-\omega _q^2}}+\frac{q^2m^2}{16 \pi  \omega _q^3 \left(m^4-\omega _q^2\right)^2}\left[3 m^4 \omega _q^2 \sqrt{m^4-\omega _q^2} \cos ^{-1}\left(\frac{\omega _q}{m^2}\right)\right.\right.\\
&\left.\left.-2 m^8 \sqrt{m^4-\omega _q^2} \cos ^{-1}\left(\frac{\omega _q}{m^2}\right)+\left(m^4-\omega _q^2\right) \left(-2 m^4 \omega _q-\pi  m^2 \omega _q^2+\omega _q^3+\pi  m^6\right)\right]\right\}+\mathcal{O}(q^4).\\
\end{split}
\label{eq.a.28}
\end{equation}

The other integrals are simpler, so we do not need to use IBP relation. After integrating over $\omega$, we can expand the result  around $q^2\to 0$ and write it in terms of functions $J_{a,b,c}$. The results are
\begin{equation}
\begin{split}
&f_a^p(\omega _q,q^2)=\frac{1}{4}\left\{2J_{-\frac{3}{2},\frac{1}{2},1}+\frac{q^2}{2D}\left[\left((3D-1) \omega_q^4 m^4\right)J_{-\frac{1}{2},\frac{5}{2},3}\right.\right.\\
&\left.\left.+\left((5D-4)\omega_q^4 m^2-(28-16 D)\omega_q^2 m^6\right)J_{-\frac{3}{2},\frac{5}{2},3}\right.\right.\\
&\left.\left.+\left(2D\omega_q^4-(140-32D)\omega_q^2 m^4+(16 D -32)m^8\right)J_{-\frac{5}{2},\frac{5}{2},3}\right.\right.\\
&\left.\left.+\left((16D-160) \omega_q^2m^2+(16 D-192)m^6\right)J_{-\frac{7}{2},\frac{5}{2},3}\right.\right.\\
&\left.\left.-80Dm^2J_{-\frac{9}{2},\frac{5}{2},3}+(64-32D)J_{-\frac{11}{2},\frac{5}{2},3}\right]\right\}+\mathcal{O}\left(q^4\right).\\
\end{split}
\end{equation}
\label{eq.a.29}
\begin{equation}
\begin{split}
&f_b^p(\omega _q,q^2)=\frac{1}{4}\left\{2J_{\frac{1}{2},-\frac{3}{2},1}+\frac{q^2}{2D}\left[\left((3-D) \omega_q^4 m^4\right)J_{\frac{3}{2},\frac{1}{2},3}\right.\right.\\
&\left.\left.+\left((4+D)\omega_q^4 m^2-(16D-36)\omega_q^2 m^6\right)J_{\frac{1}{2},\frac{1}{2},3}\right.\right.\\
&\left.\left.+\left(2D\omega_q^4-(32D-52)\omega_q^2 m^4+(160 -48D)m^8\right)J_{-\frac{1}{2},\frac{1}{2},3}\right.\right.\\
&\left.\left.+\left(-(32+16D) \omega_q^2m^2+(576-176D)m^6\right)J_{-\frac{3}{2},\frac{1}{2},3}\right.\right.\\
&\left.\left.+\left(-48\omega_q^2+(736-240D)m^4\right)J_{-\frac{5}{2},\frac{1}{2},3}\right.\right.\\
&\left.\left.+((384-144D)m^2)J_{-\frac{7}{2},\frac{1}{2},3}+(64-32D)J_{-\frac{9}{2},\frac{1}{2},3}\right]\right\}+\mathcal{O}\left(q^4\right).\\
\end{split}
\label{eq.a.30}
\end{equation}
\begin{equation}
\begin{split}
&f_d^p(\omega _q,q^2)=\frac{1}{2}\left\{2J_{-\frac{1}{2},-\frac{1}{2},1}+\frac{q^2}{2D}\left[\left((3-D) \omega_q^4 m^4\right)J_{\frac{1}{2},\frac{3}{2},3}\right.\right.\\
&\left.\left.+\left((4+D)\omega_q^4 m^2-(16D-36)\omega_q^2 m^6\right)J_{-\frac{1}{2},\frac{3}{2},3}\right.\right.\\
&\left.\left.+\left(2D\omega_q^4-(32D-52)\omega_q^2 m^4+(160 -48D)m^8\right)J_{-\frac{3}{2},\frac{3}{2},3}\right.\right.\\
&\left.\left.+\left(-(32+16D) \omega_q^2m^2+(576-176D)m^6\right)J_{-\frac{5}{2},\frac{3}{2},3}\right.\right.\\
&\left.\left.+\left(-48\omega_q^2+(736-240D)m^4\right)J_{-\frac{7}{2},\frac{3}{2},3}\right.\right.\\
&\left.\left.+((384-144D)m^2)J_{-\frac{9}{2},\frac{3}{2},3}+(64-32D)J_{-\frac{11}{2},\frac{3}{2},3}\right]\right\}+\mathcal{O}\left(q^4\right).\\
\end{split}
\label{eq.a.31}
\end{equation}
\begin{equation}
\begin{split}
&f_f^p(\omega _q,q^2)=-\frac{\omega_q}{2}\left\{J_{-\frac{1}{2},\frac{1}{2},1}+\frac{q^2}{D}\left[D\omega_q^4 m^2J_{\frac{1}{2},\frac{3}{2},3}
+\left(D\omega_q^4 -(7-6D)\omega_q^2 m^4\right)J_{-\frac{1}{2},\frac{3}{2},3}\right.\right.\\
&\left.\left.+\left((10D-28)\omega_q^2m^2+(8D-12)m^6\right)J_{-\frac{3}{2},\frac{3}{2},3}
             +\left((4D-20)\omega_q^2+(16D-60)m^4\right)J_{-\frac{5}{2},\frac{3}{2},3}\right.\right.\\
&\left.\left.+(8D-64)m^2J_{-\frac{7}{2},\frac{3}{2},3}-16J_{-\frac{9}{2},\frac{3}{2},3}\right]\right\}+\mathcal{O}\left(q^4\right).\\
\end{split}
\label{eq.a.32}
\end{equation}
\begin{equation}
\begin{split}
&f_g^p(\omega _q,q^2)=-\frac{\omega_q}{2}\left\{J_{\frac{1}{2},-\frac{1}{2},1}+\frac{q^2}{D}\left[(D\omega_q^4+(1-2D)\omega_q^2m^4)J_{\frac{1}{2},\frac{1}{2},3}\right.\right.\\
&\left.\left.+\left((2D-12)\omega_q^2 m^4+(20-8D)m^6\right)J_{-\frac{1}{2},\frac{1}{2},3}
             +\left((4D-20)\omega_q^2+(36-16D)m^4\right)J_{-\frac{3}{2},\frac{1}{2},3}\right.\right.\\
&\left.\left.-8Dm^2J_{-\frac{5}{2},\frac{1}{2},3}-16J_{-\frac{7}{2},\frac{1}{2},3}\right]\right\}+\mathcal{O}\left(q^4\right).\\
\end{split}
\label{eq.a.33}
\end{equation}

In $D=2-\epsilon$ dimensions, we find
\begin{equation}
\begin{split}
&f_a^p(\omega _q,q^2)=\frac{1}{4}\left\{\frac{N_{\rm UV}}{8\pi}+\frac{1}{8\omega_q}\left(\frac{\left(2 m^4-\omega _q^2\right) \cos ^{-1}\left(\frac{\omega _q}{m^2}\right)}{\pi  \sqrt{m^4-\omega _q^2}}-m^2\right)\right.\\
&\left.+\frac{q^2m^2}{16\pi\omega_q^3\left(m^4-\omega_q^2\right)^{\frac{3}{2}}}\left[-9 \omega _q^3 \sqrt{m^4-\omega _q^2}+10 m^4 \omega _q \sqrt{m^4-\omega _q^2}+5 \pi  m^2 \omega _q^2 \sqrt{m^4-\omega _q^2}\right.\right.\\
&\left.\left.+\left(-15 m^4 \omega _q^2+4 \omega _q^4+10 m^8\right) \cos ^{-1}\left(\frac{\omega _q}{m^2}\right)-5 \pi  m^6 \sqrt{m^4-\omega _q^2}\right]\right\}+\mathcal{O}\left(q^4\right).\\
\end{split}
\label{eq.a.34}
\end{equation}
\begin{equation}
\begin{split}
&f_b^p(\omega _q,q^2)=\frac{1}{4}\left\{\frac{N_{\rm UV}}{8\pi}+\frac{1}{8\omega_q}\left(\frac{\left(2 m^4-\omega _q^2\right) \cos ^{-1}\left(\frac{\omega _q}{m^2}\right)}{\pi  \sqrt{m^4-\omega _q^2}}+m^2\right)-\frac{q^2m^2}{16\pi\omega_q^3\left(m^4-\omega_q^2\right)^2}\left[2 m^8 \omega _q\right.\right.\\
&\left.\left.-5 m^4 \omega _q^3+\omega _q^4 \left(4 \sqrt{m^4-\omega _q^2} \cos ^{-1}\left(\frac{\omega _q}{m^2}\right)+\pi  m^2\right)-m^4 \omega _q^2 \left(5 \sqrt{m^4-\omega _q^2} \cos ^{-1}\left(\frac{\omega _q}{m^2}\right)+2 \pi  m^2\right)\right.\right.\\
&\left.\left.+m^8 \left(2 \sqrt{m^4-\omega _q^2} \cos ^{-1}\left(\frac{\omega _q}{m^2}\right)+\pi  m^2\right)+3 \omega _q^5\right]\right\}+\mathcal{O}\left(q^4\right).\\
\end{split}
\label{eq.a.35}
\end{equation}
\begin{equation}
\begin{split}
&f_d^p(\omega _q,q^2)=\frac{1}{2}\left\{\frac{N_{\rm UV}}{8\pi}-\frac{\omega _q \cos ^{-1}\left(\frac{\omega _q}{m^2}\right)}{8 \pi  \sqrt{m^4-\omega _q^2}}+\frac{q^2m^2}{16\pi\omega_q^3\left(m^4-\omega_q^2\right)^2}\left[-m^4 \omega _q^2 \sqrt{m^4-\omega _q^2} \cos ^{-1}\left(\frac{\omega _q}{m^2}\right)\right.\right.\\
&\left.\left.+2 m^8 \sqrt{m^4-\omega _q^2} \cos ^{-1}\left(\frac{\omega _q}{m^2}\right)-\left(m^4-\omega _q^2\right) \left(-2 m^4 \omega _q-\pi  m^2 \omega _q^2+3 \omega _q^3+\pi  m^6\right)\right]\right\}+\mathcal{O}\left(q^4\right).\\
\end{split}
\label{eq.a.36}
\end{equation}
\begin{equation}
\begin{split}
&f_f^p(\omega _q,q^2)=-\frac{\omega_q}{2}\left\{\frac{\pi -\frac{2 m^2 \cos ^{-1}\left(\frac{\omega _q}{m^2}\right)}{\sqrt{m^4-\omega _q^2}}}{16 \pi  \omega _q}-\frac{q^2}{16\pi\omega_q^3\left(m^4-\omega_q^2\right)^{\frac{3}{2}}}\left[6 m^4 \omega _q \sqrt{m^4-\omega _q^2}\right.\right.\\
&\left.\left.+3 \pi  m^2 \omega _q^2 \sqrt{m^4-\omega _q^2}-5 \omega _q^3 \sqrt{m^4-\omega _q^2}+\left(-11 m^4 \omega _q^2+4 \omega _q^4+6 m^8\right) \cos ^{-1}\left(\frac{\omega _q}{m^2}\right)\right.\right.\\
&\left.\left.-3 \pi  m^6 \sqrt{m^4-\omega _q^2}\right]\right\}+\mathcal{O}\left(q^4\right).\\
\end{split}
\label{eq.a.37}
\end{equation}
\begin{equation}
\begin{split}
&f_g^p(\omega _q,q^2)=-\frac{\omega_q}{2}\left\{\frac{\frac{2 m^2 \cos ^{-1}\left(\frac{\omega _q}{m^2}\right)}{\sqrt{m^4-\omega _q^2}}+\pi }{16 \pi  \omega _q}-\frac{q^2}{16\pi\omega_q^3\left(m^4-\omega_q^2\right)^{\frac{3}{2}}}\left[4 m^4 \omega _q \sqrt{m^4-\omega _q^2}\right.\right.\\
&\left.\left.+\pi  m^2 \omega _q^2 \sqrt{m^4-\omega _q^2}-5 \omega _q^3 \sqrt{m^4-\omega _q^2}+\left(-7 m^4 \omega _q^2+4 \omega _q^4+4 m^8\right) \cos ^{-1}\left(\frac{\omega _q}{m^2}\right)\right.\right.\\
&\left.\left.-\pi  m^6 \sqrt{m^4-\omega _q^2}\right]\right\}+\mathcal{O}\left(q^4\right).\\
\end{split}
\label{eq.a.38}
\end{equation}

\end{document}